\documentclass[referee,a4paper,12pt,traditabstract]{swsc} 

\usepackage{amssymb,amsthm}
\usepackage{amsmath}
\usepackage{gensymb}

\usepackage{graphicx}
\usepackage{txfonts}
\usepackage{epstopdf}
\usepackage{lineno}
\usepackage[authoryear,round]{natbib}
\usepackage[backref]{hyperref}
\usepackage{url}

\usepackage{float}
\usepackage{deluxetable}
\usepackage{nomencl}
\usepackage[labelfont=bf,textfont=it,labelsep=quad]{caption}
\usepackage{lscape}
\usepackage{longtable}
\usepackage{enumitem}
\usepackage{textcomp}

\newcommand\Tstrut{\rule{0pt}{3ex}}         
\newcommand\Bstrut{\rule[-1.5ex]{0pt}{0pt}}   

\hypersetup{colorlinks=true,citecolor=cyan,urlcolor=cyan,linkcolor=blue}

\begin{document}


   \title{\textbf{Solar signatures and eruption mechanism of the 2010~August~14 CME}}
   
   \titlerunning{The 2010~August~14 CME}

   \authorrunning{D'Huys et al.}

   \author{E. D'Huys\inst{1, 2}\thanks{\email{\href{mailto:elke.dhuys@observatory.be}{elke.dhuys@observatory.be}}}
          \and
          D.~B. Seaton\inst{1, 3, 4}
          \and
          A. De Groof\inst{5}
          \and
          D. Berghmans\inst{1}
          \and
          S. Poedts\inst{2}
          }

   \institute{Solar-Terrestrial Center for Excellence\\
              Royal Observatory of Belgium\\
              Solar Influences Data Analysis Center\\
              Avenue Circulaire -3- Ringlaan, 1180 Brussels, Belgium\\
         \and
             Centre for mathematical Plasma-Astrophysics\\
             Katholieke Universiteit Leuven \\
             Leuven, Belgium\\
         \and
             Cooperative Institute for Research in Environmental Sciences\\	
             University of Colorado at Boulder\\
             Boulder, Colorado, U.S.A.\\
         \and
             NOAA National Centers for Environmental Information\\
             Boulder, Colorado, U.S.A.\\
         \and
             European Space Agency / ESAC\\
             Camino Bajo del Castillo s/n\\
             E-28692 Villanueva de la Cañada, Madrid, Spain \\
             }


  \abstract
   {On 2010 August 14, a wide-angled coronal mass ejection (CME) was observed. This solar eruption originated from a destabilized filament that connected two active regions and the unwinding of this filament gave the eruption an untwisting motion that drew the attention of many observers. In addition to the erupting filament and the associated CME, several other low-coronal signatures that typically indicate the occurrence of a solar eruption were associated to this event. However, contrary to what is expected, the fast CME ($\mathrm{v}>900~\mathrm{km}~\mathrm{s}^{-1}$) was accompanied by only a weak C4.4 flare.
  
   We investigate the various eruption signatures that were observed for this event and focus on the kinematic evolution of the filament in order to determine its eruption mechanism. Had this solar eruption occurred just a few days earlier, it could have been a significant event for space weather. The risk to underestimate the strength of this eruption based solely on the C4.4 flare illustrates the need to include all eruption signatures in event analyses in order to obtain a complete picture of a solar eruption and assess its possible space weather impact.  
} 
   
   \keywords{solar corona --
             solar eruptions --
             solar flares --
             coronal mass ejections --
             solar energetic particles
               }

   \maketitle
\section{Introduction}

\subsection{The 2010 August 14 CME}

An unusual filament eruption occurred on the north-west solar limb on 2010 August 14. It was one of the first strong eruptions of solar cycle 24, which had just started to rise in activity at the time. Therefore, space weather forecasters carefully tracked and analyzed the event. Although this eruption was associated with a fast coronal mass ejection (CME), the flare that accompanied it was atypically weak. However, the eruption was also the source of the first proton event recorded in nearly four years. Interestingly, the erupting filament exhibited a notable unwinding motion as it was accelerated away from the Sun. It is this twist in the unraveling filament that first attracts the attention when studying the EUV observations of the eruption. Despite the weakness of the associated C4.4 flare, it became immediately clear from the coronagraph data that this event would be significant for space weather: a wide CME was launched into interplanetary space with a very high velocity (over $900~\mathrm{km}~\mathrm{s}^{-1}$, see Section~\ref{recon_section}). 

From a forecasting perspective, whenever such a solar eruption is observed, the first task is to gather as many observations as possible. These observations may be in extreme ultraviolet (EUV), white light (WL), from radio instruments and coronagraphs; as well as measurements from particle instruments. Additionally, forecasters rely on the output of various automated detection tools such as CACTus \citep[CMEs;][]{Robbrecht2009_cactus} and Solar Demon \citep[solar flares, dimmings and EUV waves;][]{Kraaikamp2015} to build a complete and coherent picture of the eruption. Often, there is limited information available to base an initial assessment on, as it takes time to gather and process all relevant data. Also in the case of this event, the forecaster on duty sent out the daily bulletin from the Belgian regional warning center (RWC) of the International Space Environment Service (ISES) with a preliminary interpretation of the eruption, stating that the CME was mostly southward directed, with a possible earthward component that could cause minor geomagnetic effects 3 days later. The next day more data had been acquired and the forecaster bulletin also reported on the associated EUV wave, radio shock and particle storm (described in detail in Section~\ref{obs}).

Solar eruptions can interact with Earth and our technology in two major ways. First there are Solar Energetic Particle events (SEPs) which are triggered when fast CME shocks accelerate charged particles in the solar atmosphere to very high velocities. A strong increase in the proton flux can have a number of space weather effects on technology in space and on Earth. For example, during a proton storm satellite electronics may be damaged and the ionization rate of the ionosphere may be locally increased, which in turn causes disturbances with High Frequency (HF) radio communication. Second, the magnetic field embedded in the CME may interact directly with the magnetic field of the Earth. The severity of a geomagnetic storm is determined by the CME speed as well as the strength and orientation of the north-south component of the CME magnetic field upon arrival: if this component is negative (southward), the magnetic field of the CME can reconnect with Earth's magnetic field and trigger a geomagnetic storm of which the strength is directly dependent on the strength of the southward magnetic field \citep{Savani2015}. Where we can measure the velocities of CMEs fairly easily with various spacecraft, we currently have no adequate way to estimate the orientation of the CME magnetic field at the time it arrives at Earth. We can make an estimate of the magnetic field structure when the CME is launched based on photospheric observations, but geometrical changes of the CME ---such as deflection and rotation--- as it moves through interplanetary space may directly influence its geomagnetic effectiveness \citep{Kay2016, Isavnin2014, Zuccarello2012, Chane2005}.

Aside from the possible space weather effects, many authors were interested in this event as a science case as well, especially because in 2010 the solar cycle had only just started to pick up in activity again. For example, \cite{Long2011} studied the kinematics and expansion rate of the associated coronal bright front (CBF, more often referred to as an \textit{EUV wave}). Using multi-wavelength radio observations, \cite{Tun2013} derived the properties of the magnetic field within the core of the CME associated with this eruption. We took a different interest in the 2010 August 14 CME and focused on what the strength of various signatures that accompanied this eruption tells us about the eruption mechanism and how this may influence the analysis of the event's space weather impact. 

\cite{Steed2011} studied the similarities between the event on 2010 August 14 and a homologous eruption that originated from the same source regions on 2010 August 7. Both events are associated with the eruption of a reverse S-shaped filament structure, a flare, a coronal dimming and an EUV wave. The coronagraphs on-board the STEREO spacecraft \citep{Howard2008} observed a halo CME in both cases. Further collaborative study \citep{Steed2012} showed that while for both events similar signs of eruption (such as the reverse S-shaped flux rope structures and EUV waves) were observed, their interplanetary evolution was rather different. These authors emphasize that minor differences between the CMEs close to the solar surface (such as small differences in velocity and propagation angle), as well as variations in the surrounding solar wind (for example, compression by a high speed solar wind stream from a coronal hole), may result in very different interplanetary propagation profiles for homologous CMEs. 
 
\subsection{Models for CME Initiation}

It is clear that the required energy to power a strong CME can only come from the magnetic field, which must be in an equilibrium state before it erupts \citep{Chen2011}. Due to the gradual evolution of the photospheric field, currents accumulate slowly in the corona and therefore the corona evolves quasi-statically, as in a sequence of force-free equilibria \citep{Aulanier2014}. The exact physical mechanism that triggers coronal mass ejections is not yet fully understood, however, the crucial ingredient is a restructuring of the magnetic field that results in a loss of equilibrium and an eruption.

Flux ropes in the solar corona are held in equilibrium by a balance of the outward forces inside the flux rope that tend to push it to expand (magnetic pressure) and the inward and downward forces from the surrounding magnetic field that restrain the flux rope (magnetic tension). Once the equilibrium between these two forces is lost, the flux rope is allowed to rise until a new equilibrium is reached. If the flux rope does not reach a new equilibrium state, it will erupt catastrophically.

Several physical processes may facilitate an eruption by eroding the equilibrium of the flux rope until it reaches a new meta-stable state or a state close to eruption. In case a new meta-stable state is reached, the flux rope can resist small perturbations and a strong trigger with significant observable features is needed to initiate an eruption. If the flux rope is close to a loss of equilibrium, however, even a minimal change in the magnetic field parameters --- for example the magnetic twist --- can directly trigger an eruption. \citep{Chen2011} 

\cite{Aulanier2014} argues that there are only two mechanisms that can actually initiate solar eruptions: the torus instability, in which the internal magnetic pressure of the flux rope becomes stronger than the magnetic tension that restrains it, and breakout \cite{Antiochos1999}, in which magnetic reconnection above the flux rope removes magnetic tension until the flux rope can no longer be restrained. In either case, the result is the same: the unbalanced outward force on the flux rope causes it to erupt, and tether-cutting reconnection behind the flux rope accelerates the nascent CME and releases the stored magnetic energy that produces a solar flare.

Other authors have described different mechanisms that might facilitate the evolution towards the loss of equilibrium that causes previously stable structures in the corona to erupt. \cite{Forbes2006} list possibilities that include flux emergence \citep{Chen1989} and flux cancellation \citep{Lin2000}, which could alter the strength of the overlying magnetic field relative to the outward pressure on the flux rope, and magnetic breakout, which leads to an overall topological change in the field that restrains the pre-eruption flux rope. Still other authors have described additional possibilities such as mass loading or drainage \citep[see, for example,][]{Seaton2011}, but in almost every case the general outline of the process is the same: loss of equilibrium facilitated by one process or another leads to an eruption. 

We refer to the various processes like those described by \cite{Forbes2006} that can lead to loss of equilibrium as \textit{CME facilitators}. These are the slow processes, driven by changes in the photosphere, that gradually modify the configuration of the surrounding magnetic fields, thereby eroding the equilibrium in which the flux rope sits. Once the flux rope equilibrium is sufficiently weakened that the onset of an eruption is close, the flux rope requires a final \textit{trigger} to cause a CME to unfold. 

The mechanisms described by \cite{Aulanier2014}, breakout and the torus instability, are two examples. In either case, the eruption is triggered when balance between the outward magnetic pressure and inward magnetic tension becomes disrupted. Other MHD instabilities have been studied as well. For example, the so-called kink instability occurs when twist in the flux tube reaches a critical value above which the whole structure becomes unstable \citep{Torok2005}, however the debate over whether kink instability can actually trigger eruptions --- rather than local disruptions --- is not settled \citep{Schmieder2013}.

On the other hand, some authors have argued that breakout is unlikely to lead to fast eruptions in at least some circumstances. Breakout also requires a multipolar flux configuration and at least one null point where the reconnection can occur to be a viable trigger for an eruption. \cite{Karpen2012} argued that breakout may require a strong flare reconnection to produce a fast eruption, meaning that it is unlikely to lead to fast CMEs associated with weak flares. 

Generally speaking, the different stages of a solar eruption can be described as follows \citep{Forbes2000, Chen2011}: a flux rope, which may or may not hold a prominence, is held in place by overlying magnetic field lines that are tied to the solar surface. Due to a magnetic reconfiguration or an instability, the flux rope is allowed to rise, stretching the restraining field and forming antiparallel magnetic field lines in its wake. As the flux rope rises further, and because of the line-tying of the flux rope to the photosphere, the antiparallel field below forms a current sheet in which reconnection may take place. This tether-cutting reconnection gradually removes the magnetic tension force of the overlying field lines and facilitates the rapid eruption of the core field into interplanetary space. We refer to this final process as the \textit{CME driver}.

The reconnection underneath the flux rope drives the eruption by converting stored magnetic energy into heat and kinetic energy, producing a CME and, possibly, a flare. In case no fast reconnection is allowed to take place in the current sheet, the driving is weak. However, even in that case, the flux rope may still erupt due to a loss of equilibrium or an ideal MHD instability and no flare, brightening or post-eruptive arcade need necessarily be observed as signatures of the eruption. This phenomenon is generally referred to as a \textit{stealth} CME \citep{Robbrecht2009_stealth, DHuys2014, Kilpua2014}.

\subsection{Outline}

While specific aspects of this eruption have been studied in detail by other authors, we here provide for the first time an overview of the wide range of solar signatures related to this event that are relevant to the analysis of this solar eruption (Section~\ref{obs}). In Section~\ref{recon_section} we make three-dimensional reconstructions of the erupting CME in order to study its kinematic properties and propagation into the interplanetary space. Section~\ref{initiation} focusses on the initiation mechanisms at work during this event. We conclude by discussing the relationship between the associated solar signatures of the eruption and its initiation mechanism. (Section~\ref{conclusion}).


\section{\textbf{Solar and Interplanetary Eruption Signatures}}\label{obs}
\begin{figure}  
\centering
\includegraphics[width=0.95\textwidth]{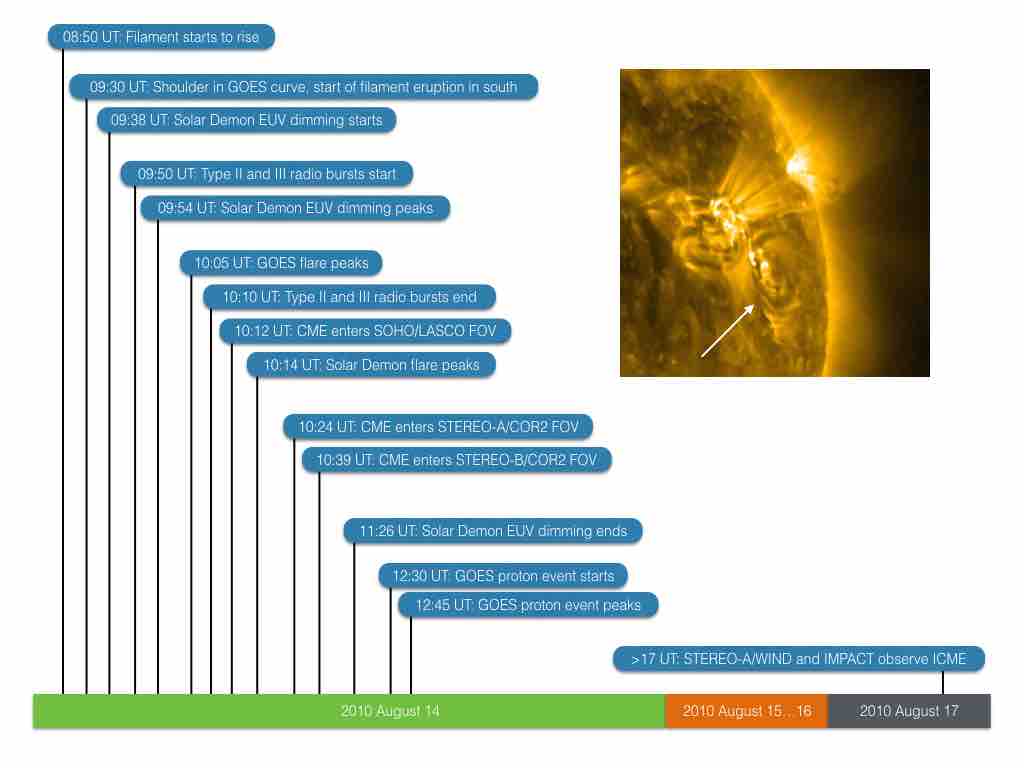}
\caption{Timeline of the most important observational signatures associated to the filament eruption on 2010 August 14. A PROBA2/SWAP image shows the filament as it erupts towards the south.}
\label{timeline} 
\end{figure}

Many solar and interplanetary signatures were associated to the eruption observed on 2010 August 14. An overview of the timing of the most important features is shown in Figure~\ref{timeline}. These signatures are discussed in detail here.

\subsection{Observations}

\subsubsection{EUV and X-ray measurements}

\begin{figure}  
\centering
\includegraphics[height=0.168\textheight]{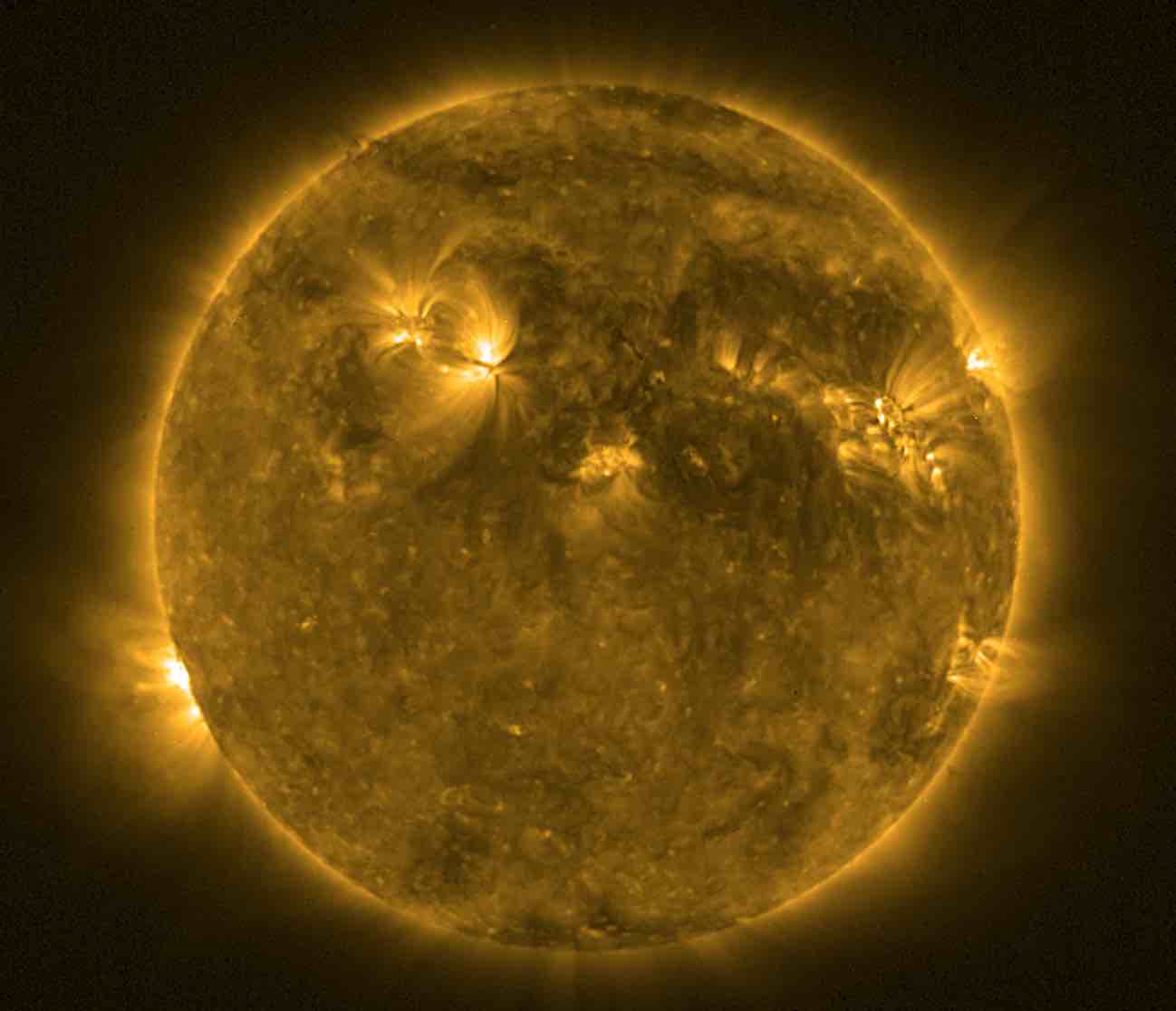}
\hspace*{-0.015\textwidth}
\includegraphics[height=0.168\textheight]{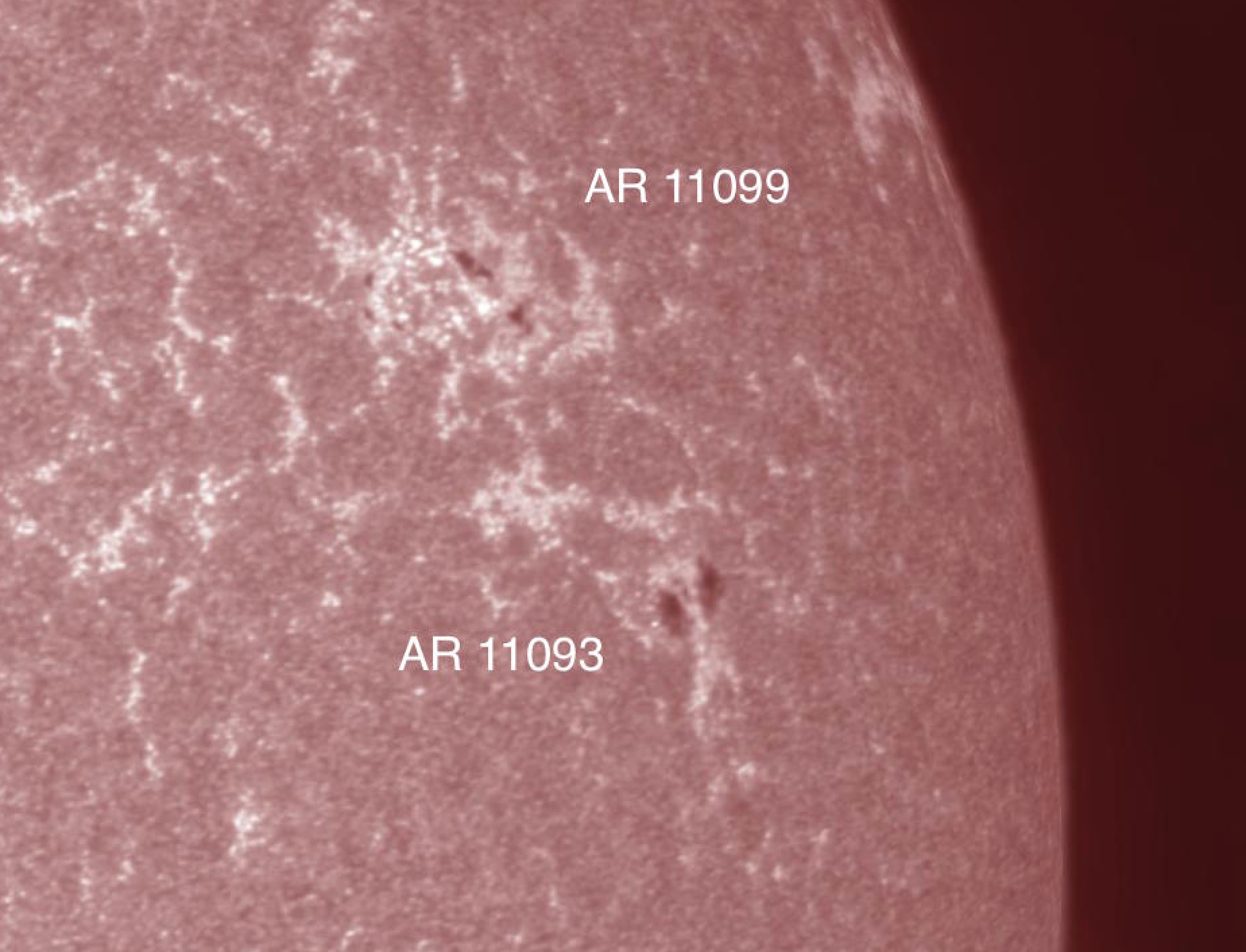}
\hspace*{-0.015\textwidth}
\includegraphics[height=0.168\textheight]{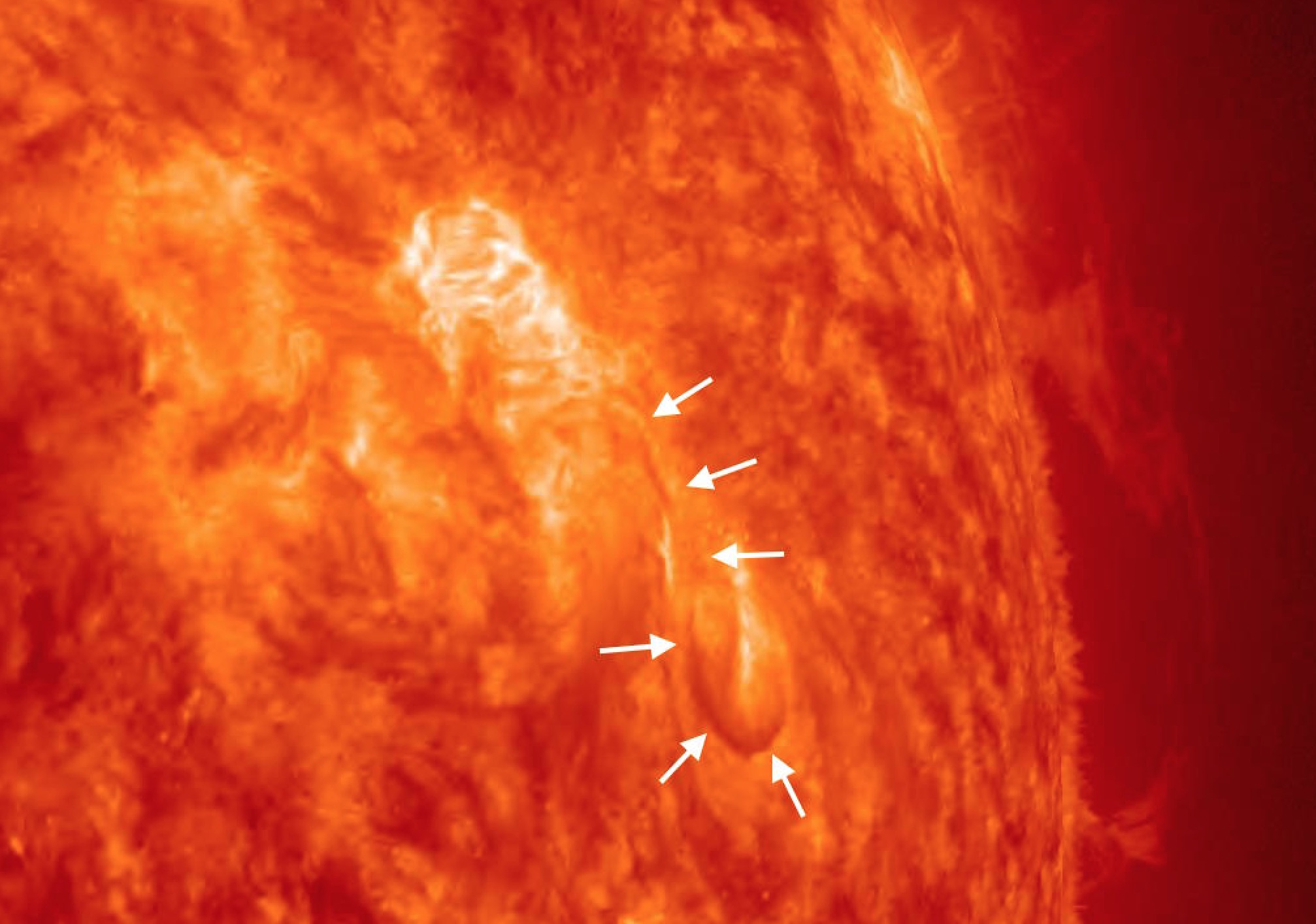}
\caption{\textbf{Left Panel:} PROBA2/SWAP 174~\AA~image of the Sun on 2010 August 14 at 04:15 UT. The eruption studied in this work occurred in the upper right quadrant. \textbf{Middle Panel:} SDO/AIA 1700~\AA~image, taken at 05:35 UT, with annotated NOAA active region numbers. \textbf{Right Panel:} SDO/AIA 304~\AA~image, taken at 08:25 UT. The white arrows indicate the location of the filament that erupted shortly after. }
\label{scenesetting} 
\end{figure}

The EUV imagers PROBA2/SWAP \citep{Seaton2013}, SDO/AIA \citep{Lemen2012} and STEREO-A/EUVI \citep{Howard2008} observed a \textit{filament eruption} on 2010 August 14 that occurred on the north-west limb of the Sun. Two separate active regions, linked by a filament, were involved: NOAA AR 11099 (Figure~\ref{scenesetting} middle panel, north) and NOAA AR 11093 (south). These active regions were classified according to their magnetic configuration as, respectively, a $\beta$ (bipolar) and an $\alpha$ (unipolar) region. PROBA2/SWAP images show these regions crackling with small-scale activity until the filament finally rises, starting around 08:50 UT.

\begin{figure}       
\centering
\includegraphics[width=0.24\textwidth,clip=]{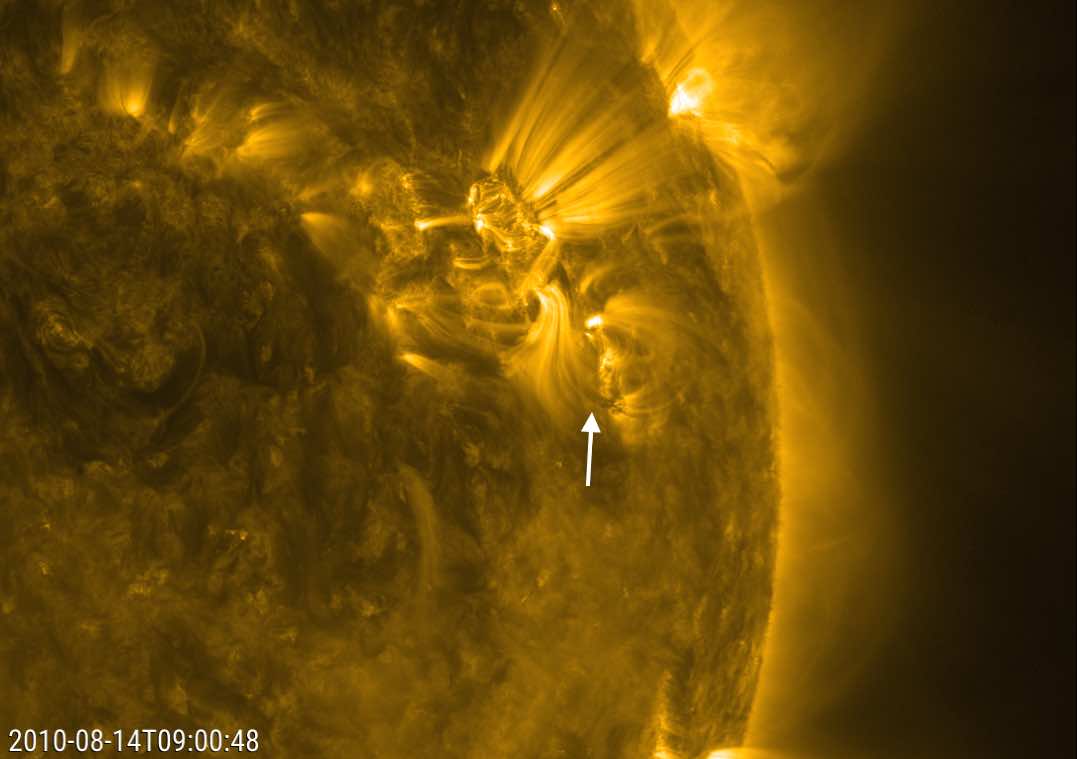}
\hspace*{-0.005\textwidth}
\includegraphics[width=0.24\textwidth,clip=]{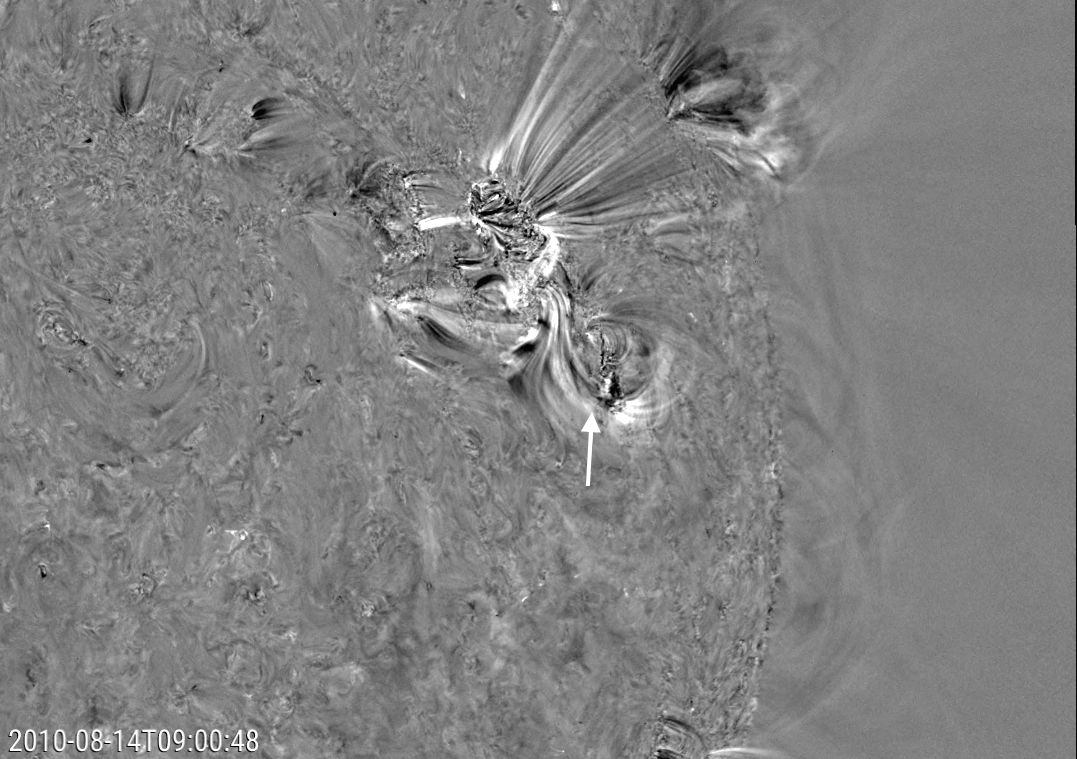}
\hspace*{-0.005\textwidth}
\includegraphics[width=0.24\textwidth,clip=]{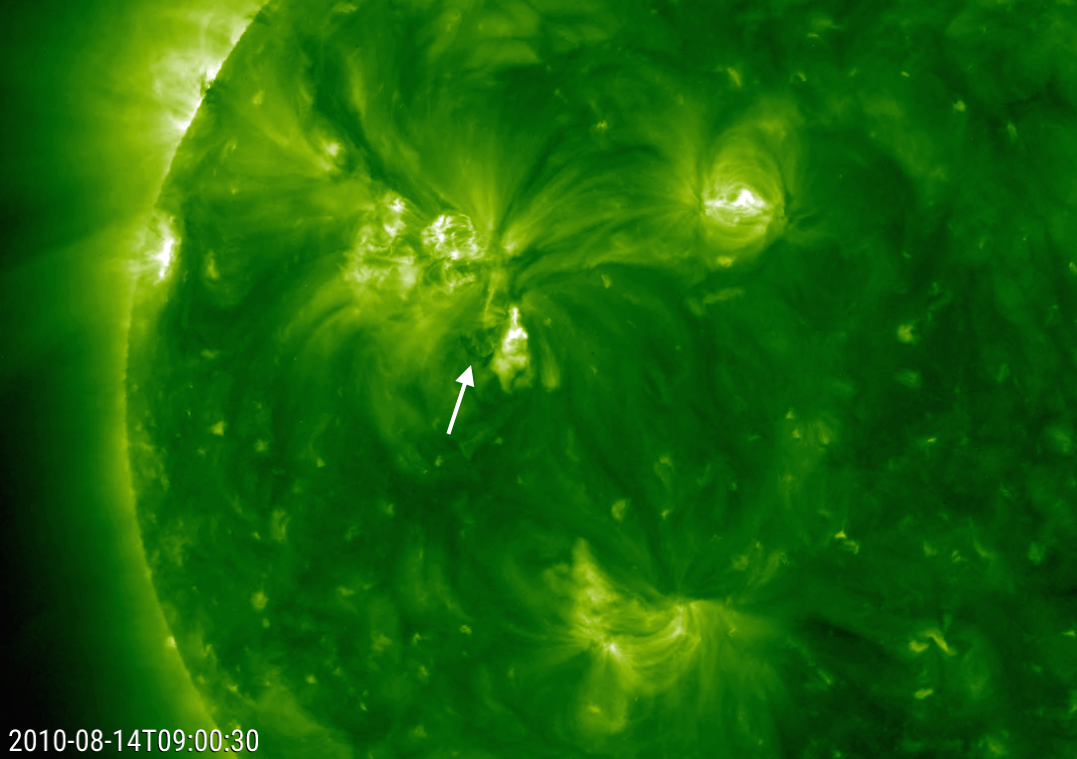}
\hspace*{-0.005\textwidth}
\includegraphics[width=0.24\textwidth,clip=]{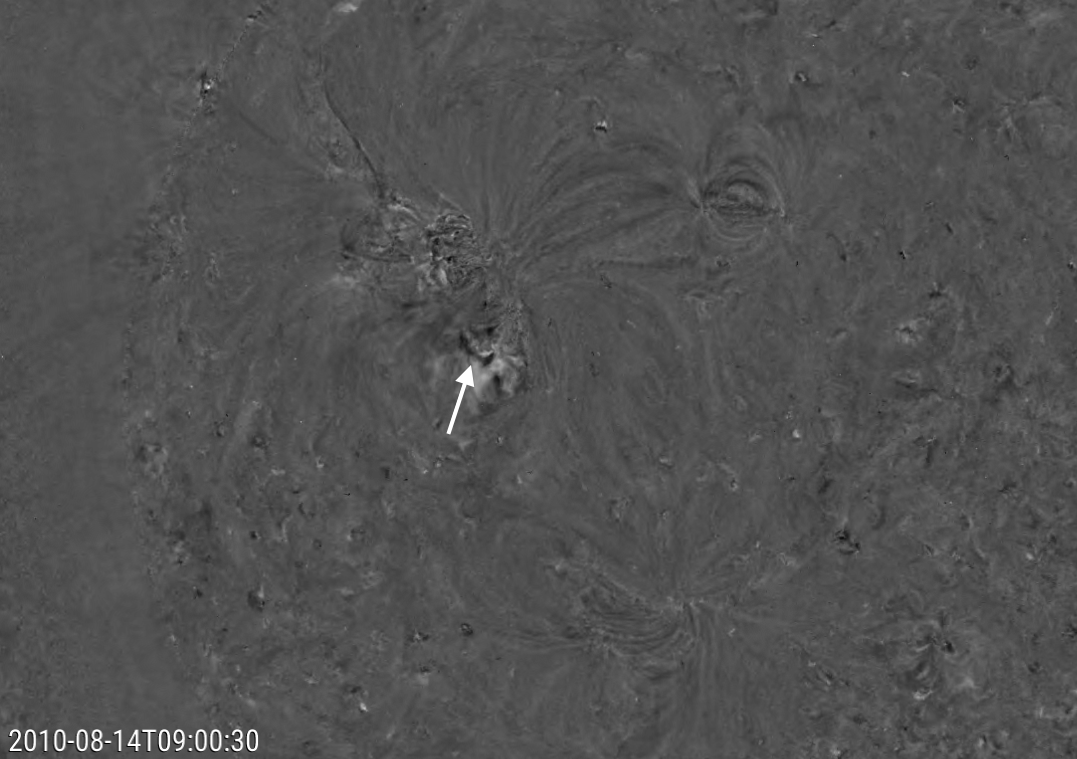}

\includegraphics[width=0.24\textwidth,clip=]{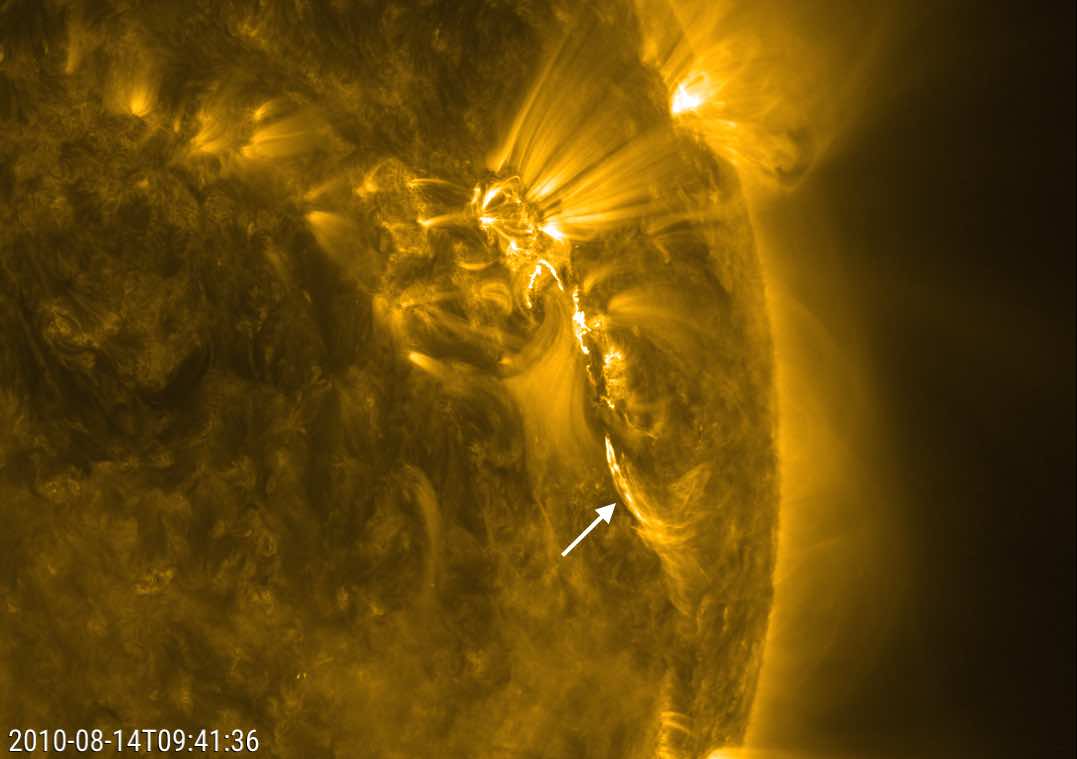}
\hspace*{-0.005\textwidth}
\includegraphics[width=0.24\textwidth,clip=]{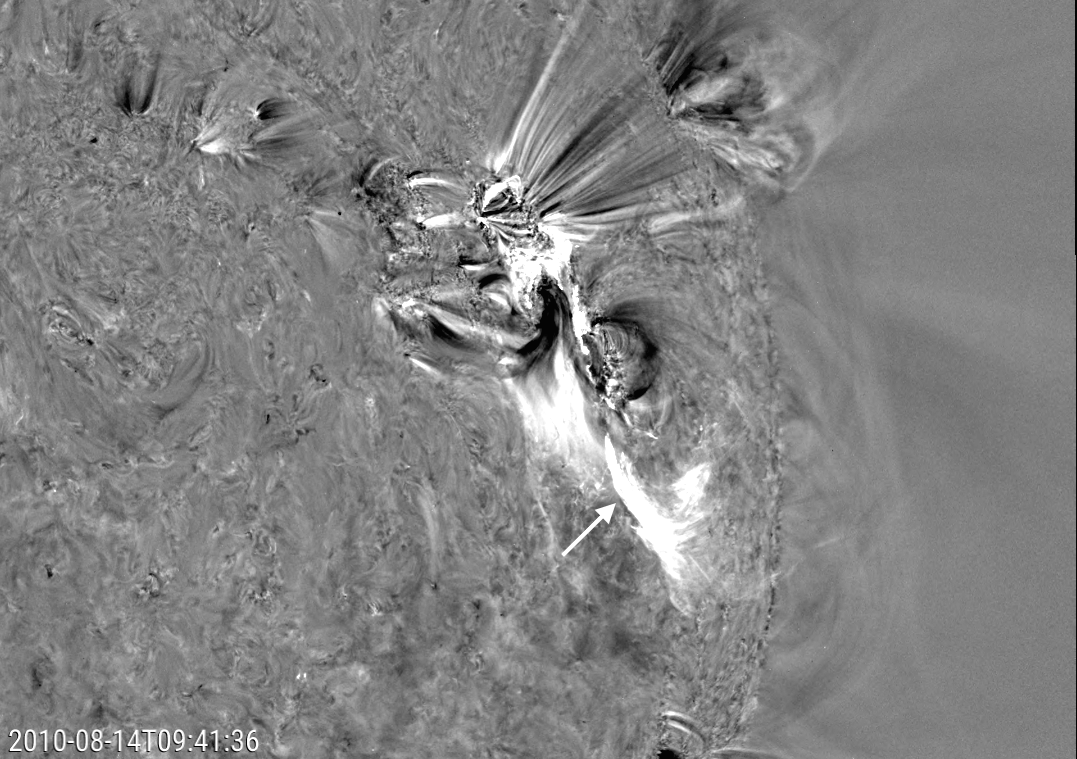}
\hspace*{-0.005\textwidth}
\includegraphics[width=0.24\textwidth,clip=]{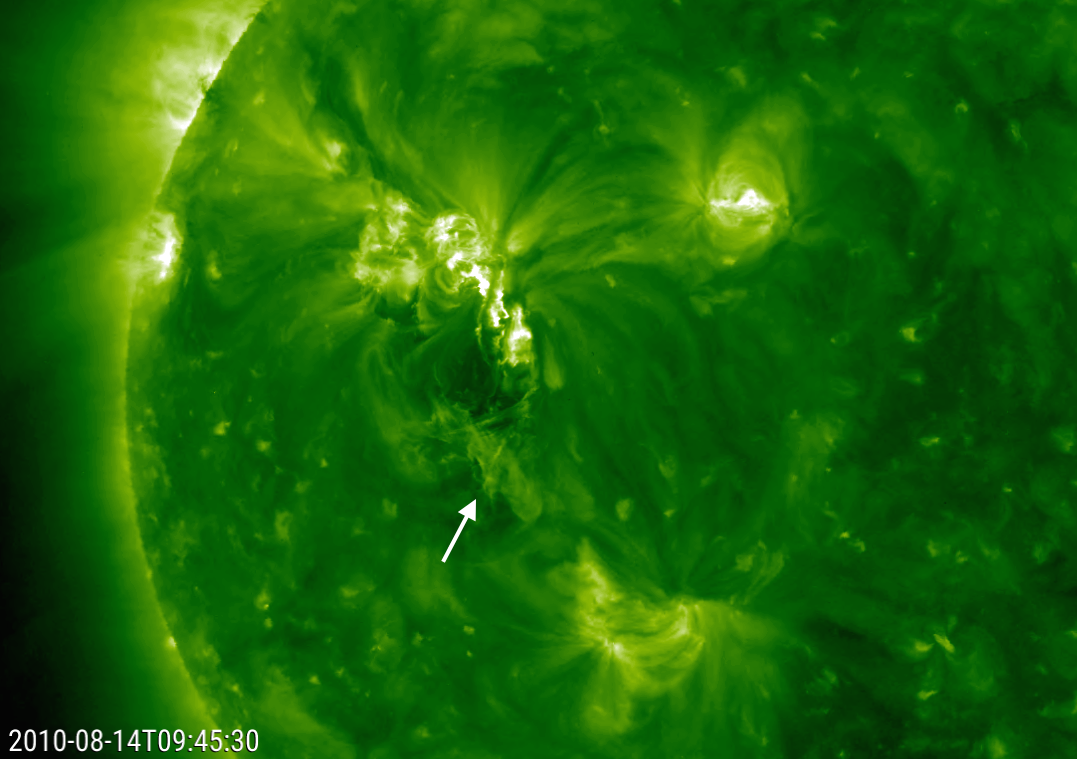}
\hspace*{-0.005\textwidth}
\includegraphics[width=0.24\textwidth,clip=]{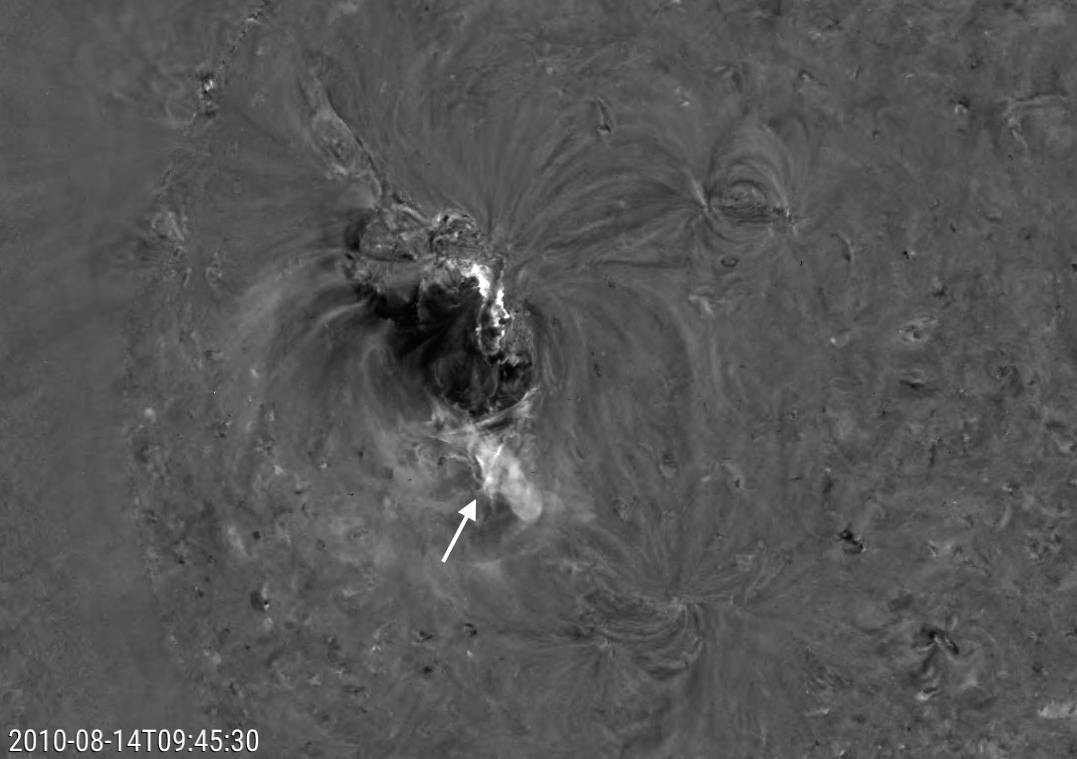}

\includegraphics[width=0.24\textwidth,clip=]{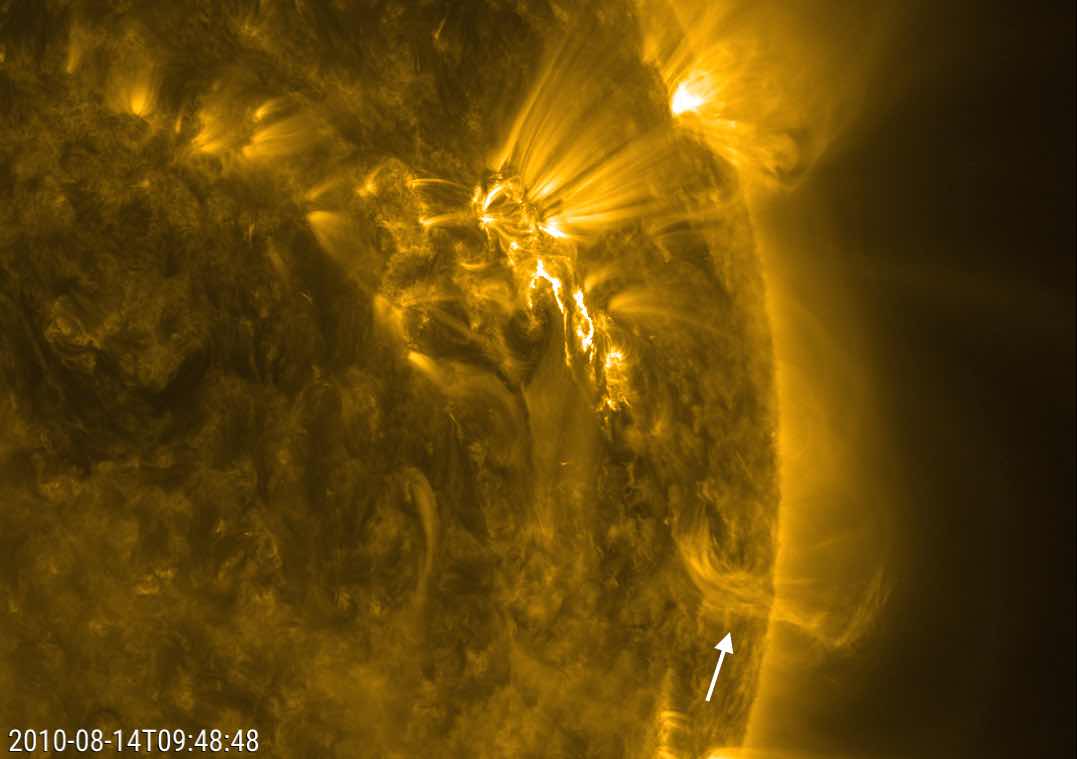}
\hspace*{-0.005\textwidth}
\includegraphics[width=0.24\textwidth,clip=]{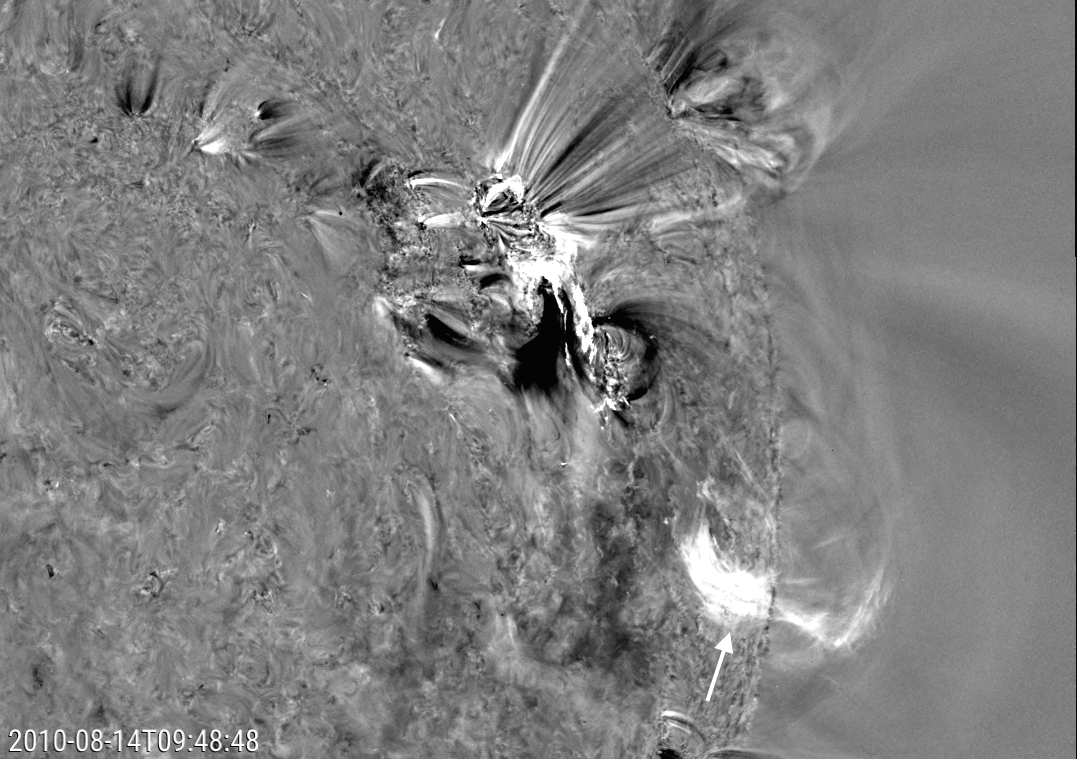}
\hspace*{-0.005\textwidth}
\includegraphics[width=0.24\textwidth,clip=]{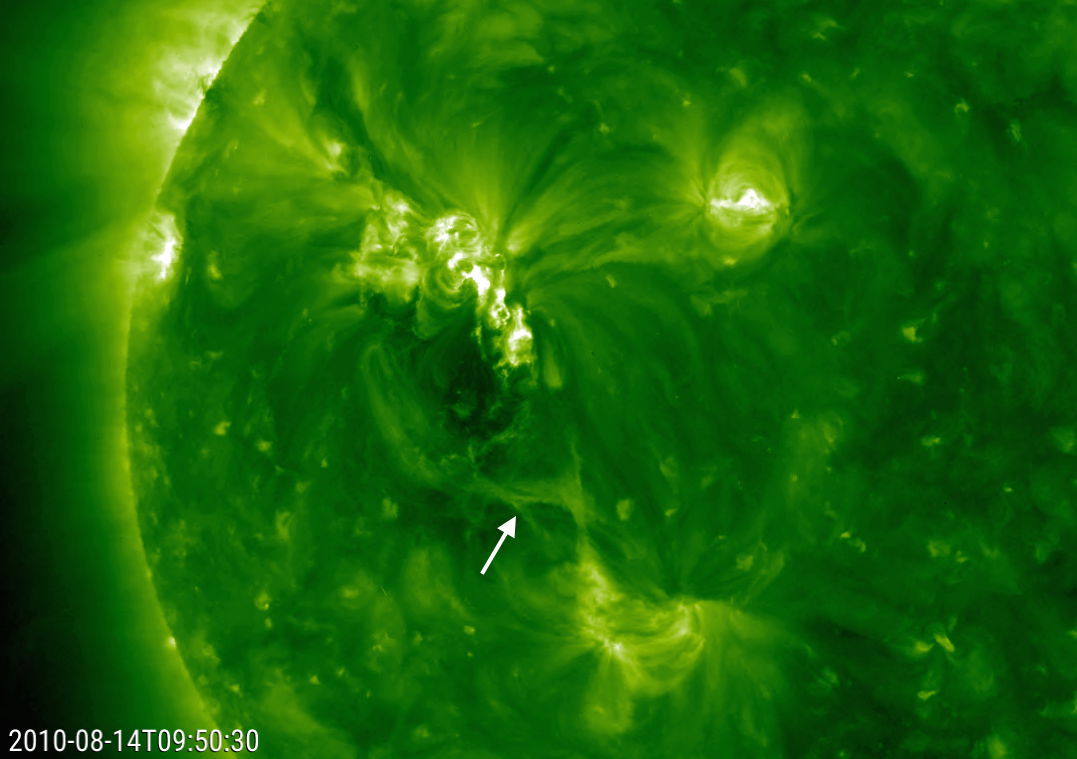}
\hspace*{-0.005\textwidth}
\includegraphics[width=0.24\textwidth,clip=]{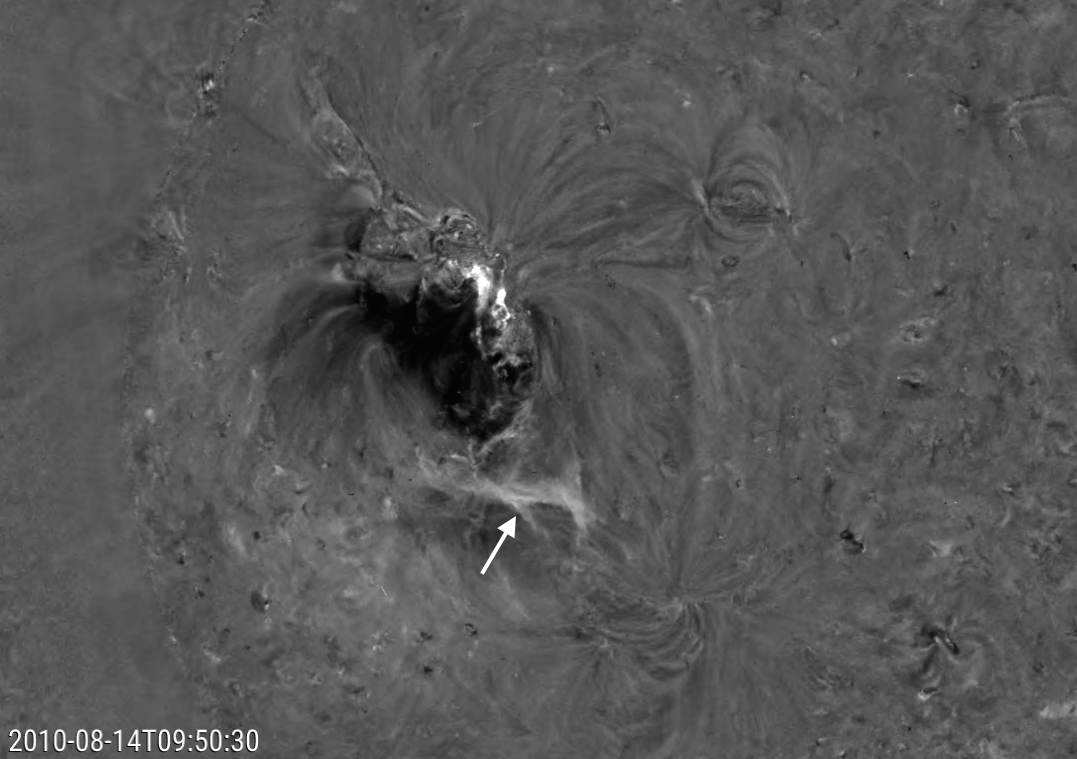}

\includegraphics[width=0.24\textwidth,clip=]{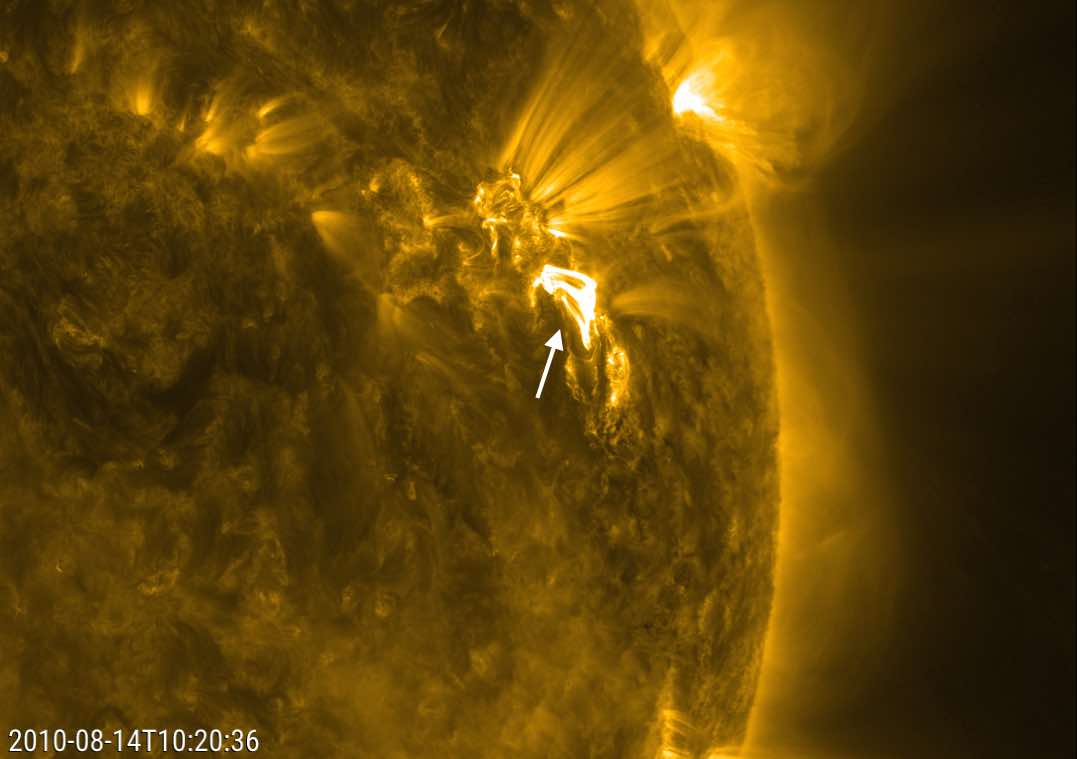}
\hspace*{-0.005\textwidth}
\includegraphics[width=0.24\textwidth,clip=]{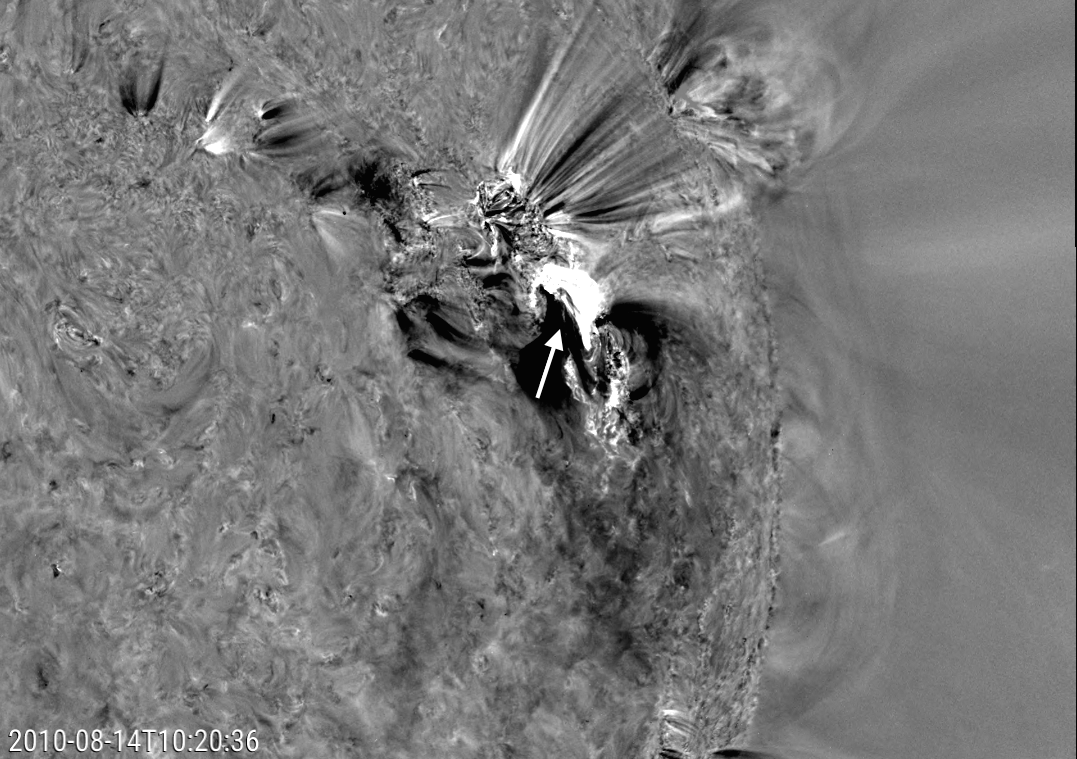}
\hspace*{-0.005\textwidth}
\includegraphics[width=0.24\textwidth,clip=]{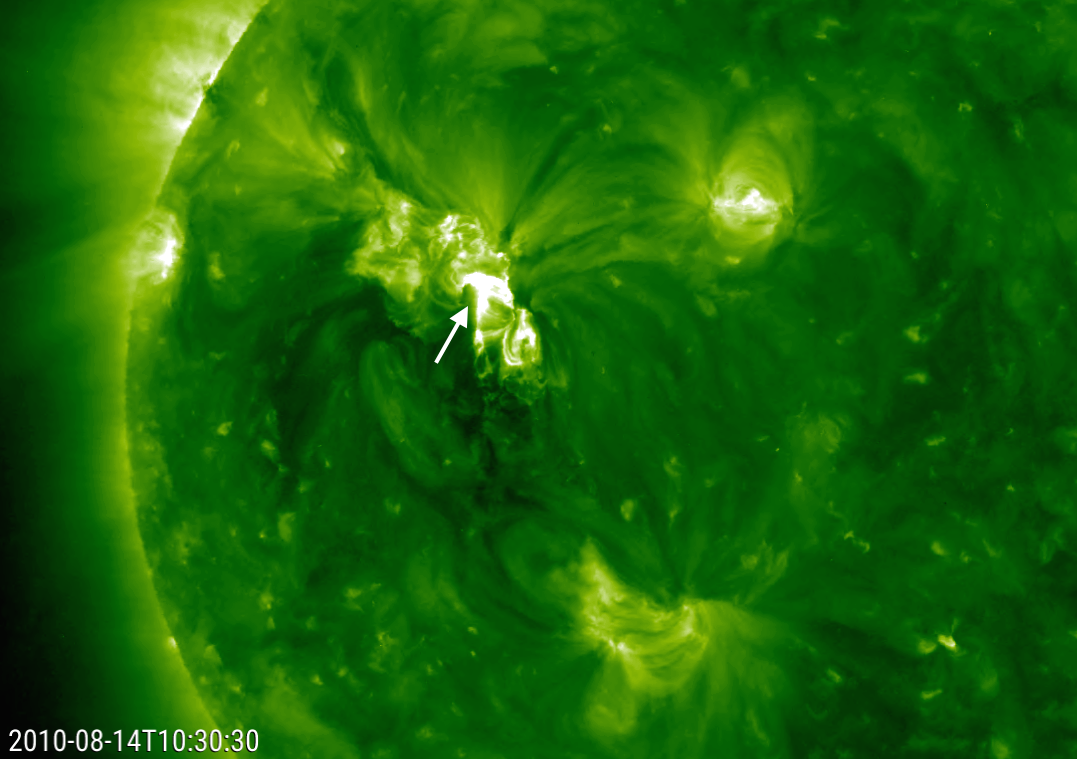}
\hspace*{-0.005\textwidth}
\includegraphics[width=0.24\textwidth,clip=]{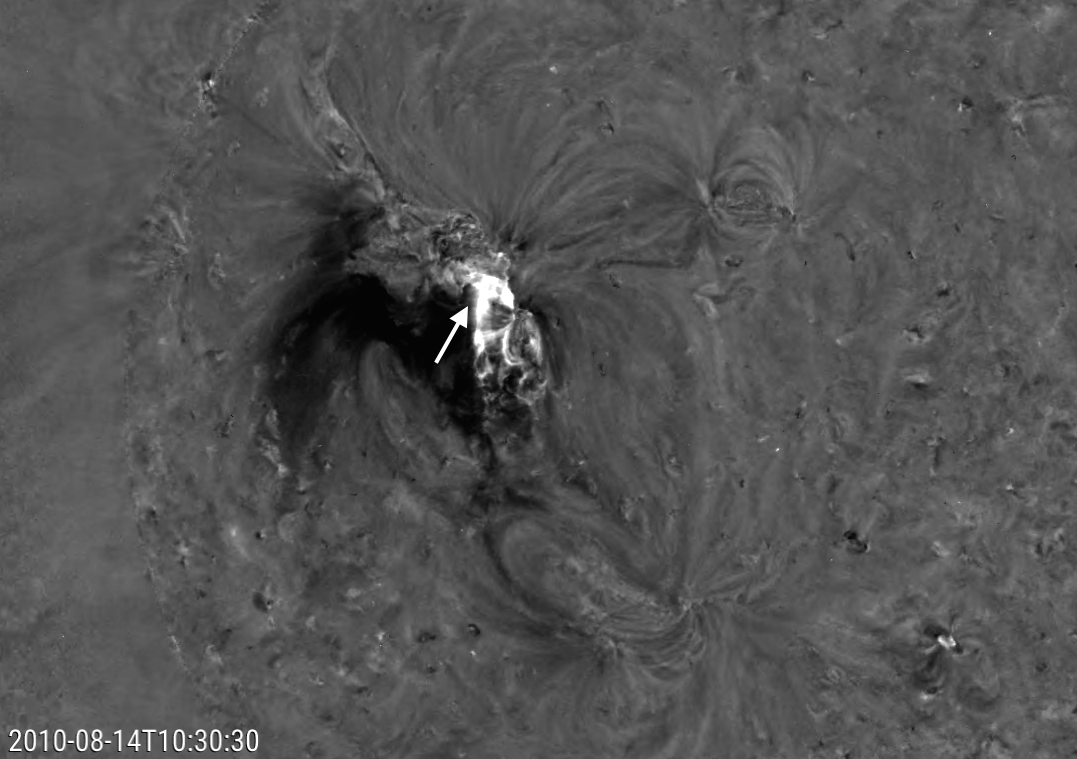}

\caption{SDO/AIA 171 \AA\ plain and base difference images (left columns) of the eruption on 2010 August 14, combined with STEREO-A EUVI 195 \AA~images (plain and base difference, right columns) taken at approximately the same time. These images show (from top to bottom and indicated by the white arrows): the rising filament, its unwinding motion as it is hurled into space, and the post-eruptive loops.}
\label{aia}
\end{figure}

Figure~\ref{aia} shows the evolution of the eruption in SDO/AIA 171~\AA\ and STEREO-A/EUVI 195 \AA\ images. The second and fourth column are base difference images, highlighting the changes in the erupting structure. Prior to the eruption, the first signs of activity are seen in the form of flickering bright points and plasma flows, mainly in the northern active region. Around 08:54~UT, SDO/AIA 171~\AA\ images show the onset of the rise of the filament (indicated in the top panels in Figure~\ref{aia} by a white arrow) that connects both active regions. The filament first starts to rise in the southern region, and subsequently drags the northern part with it until equilibrium is lost. Then the southern part of the filament erupts violently and the northern part is trapped in the corona (panels on second and third row in Figure~\ref{aia}). As the filament unwinds the erupting plasma is hurled into space with an untwisting motion. Additional discussion of the CME trajectory and onset appears below in Section~\ref{recon_section}. 

\begin{figure}  
\centering
\includegraphics[height=0.3\textheight]{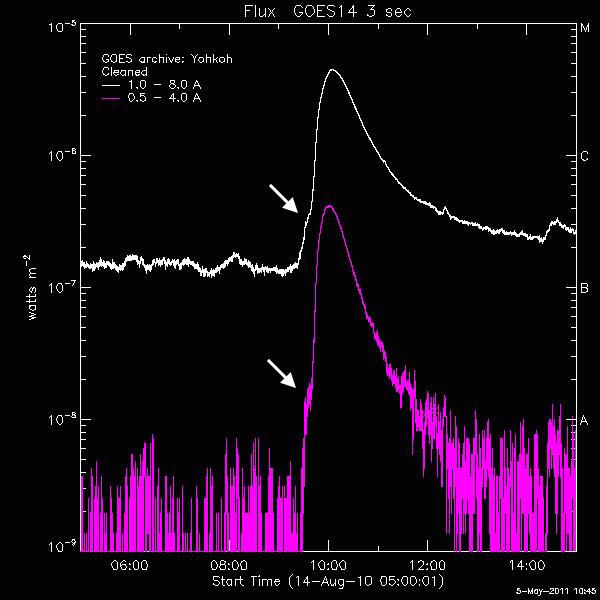}
\hspace*{-0.015\textwidth}
\includegraphics[height=0.3\textheight]{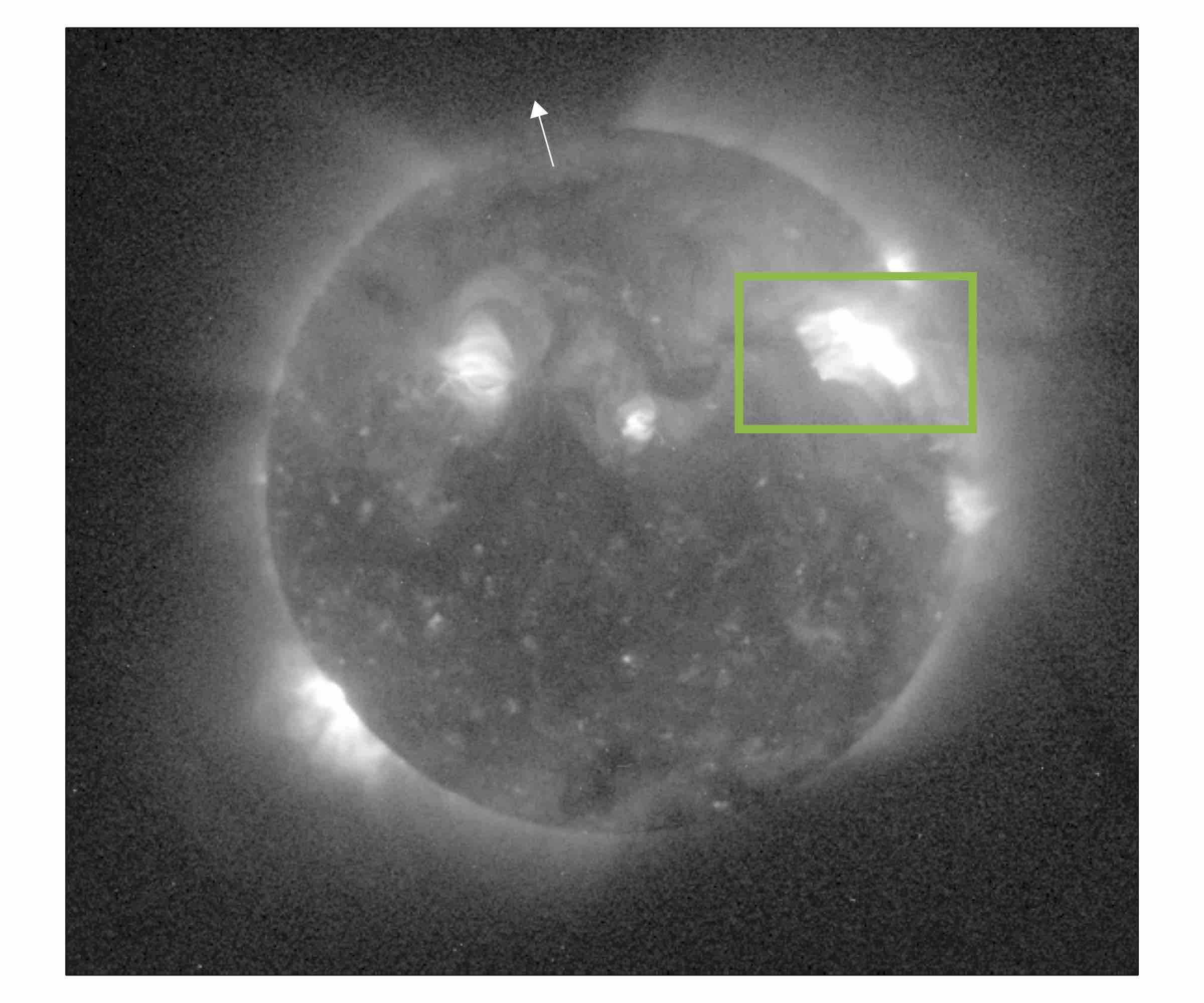}
\caption{\textbf{Left Panel:} The GOES X-ray flux measurements on 2010 August 14 show a C4.4 flare at the time of the eruption. The arrows indicate a shoulder in the rising phase around 09:30 UT, which corresponds in time to the start of the filament eruption in the southern region. \textbf{Right panel:} The associated flare imaged by the Solar X-Ray Imager on the GOES-14 spacecraft at 09:49 UT. The flaring region is bounded by the green box. (The white arrow indicates solar north.)}
\label{flare_goes} 
\end{figure}  

Both active regions showed weak flaring activity up to B-level in the days before and after the eruption. The event itself was measured by GOES/XRS \citep{Hanser1996} as a moderate C4.4 \textit{flare} with a peak time of 10:05~UT (left panel in Figure~\ref{flare_goes}). The rising phase of the flare in the GOES flux curve contains a shoulder (indicated by the arrow in the left panel of Fig.~\ref{flare_goes}) around 09:30~UT which appears to correspond to the start of the filament eruption in the southern region. However, the flare peak occurs only after the northern part of the filament erupts as well. GOES measured the start and end time of this flare as 09:38~UT and 10:31~UT, respectively. 

The \textit{post-eruptive loop} system is clearly visible in the bottom panels in Figure~\ref{aia}. These post-flare loops remain visible in EUV observations for several hours after the eruption. Note that the images in Figure~\ref{aia} also show a dark region, a transient coronal hole or \textit{dimming}, to the left of the eruption site, indicating the evacuation of coronal plasma in that region. 

\begin{figure}  
\centering
\includegraphics[height=0.35\textheight]{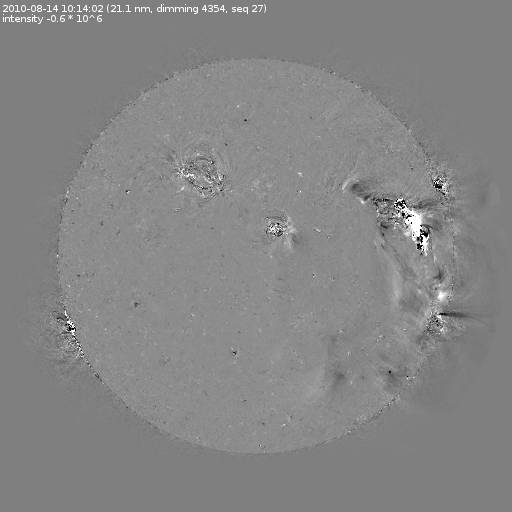}
\caption{Dimming associated with the eruption on 2010 August 14 and detected by Solar Demon in SDO/AIA 211~\AA~images. Image taken from \url{http://solardemon.oma.be/}.}
\label{dimming} 
\end{figure}

Solar Demon identifies a flare, dimming and EUV wave associated with this eruption The flare is classified as C7 by Solar Demon with peak time at 10:14~UT and it lasts from 09:44~UT to 11:24~UT, spanning 51 consecutive SDO/AIA 94~\AA~images. (Solar Demon uses a cadence of 2 minutes for flare detections.) This discrepancy in flare classification and duration can be explained by the different wavelengths of the observations used to measure these quantities by Solar Demon (EUV) and GOES (X-ray). Figure~\ref{dimming} shows the Solar Demon detection of the EUV dimming associated with this eruption in running difference images based on the SDO/AIA 211~\AA~observations. 

\begin{figure}       
\centering
\includegraphics[width=0.65\textwidth]{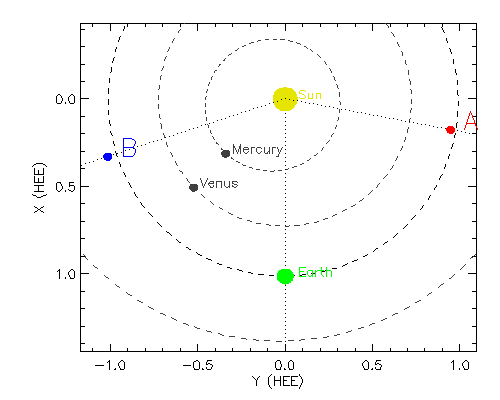}
\caption{Position of the STEREO spacecraft at 09:00 UT on 2010 August 14, with respect to the Sun and Earth. The positions of planets Venus and Mercury are indicated as well.}
\label{stereoposition}
\end{figure}

\subsubsection{White Light Observations}

This filament eruption was associated with \textit{a coronal mass ejection} observed by the coronagraphs on-board SOHO and both STEREO spacecraft as a halo CME. The location of the STEREO spacecraft at the time of the eruption and with respect to planets Earth, Venus and Mercury is illustrated in Figure~\ref{stereoposition}. The separation angle between STEREO-B and Earth was $72\degree$ on 2010 August 14, while between Earth and STEREO-A it had increased to $79.5\degree$. The SOHO and STEREO spacecraft observed this CME from three distinctly different viewpoints, providing us with a maximum of information on the angular extent, direction and velocity of this CME. 

\begin{figure}
\centering
\includegraphics[width=0.33\textwidth,clip=]{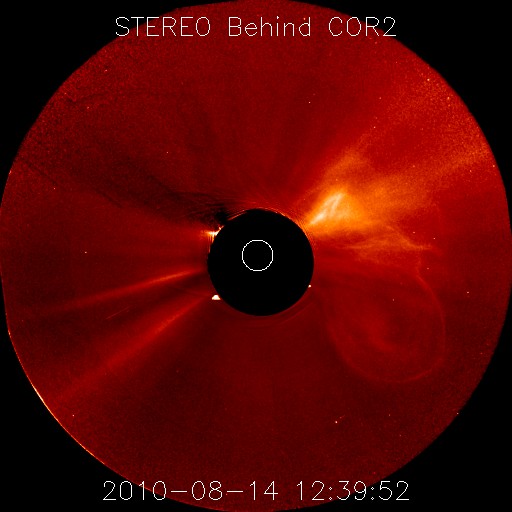} 
\hspace*{-0.015\textwidth}
\includegraphics[width=0.33\textwidth,clip=]{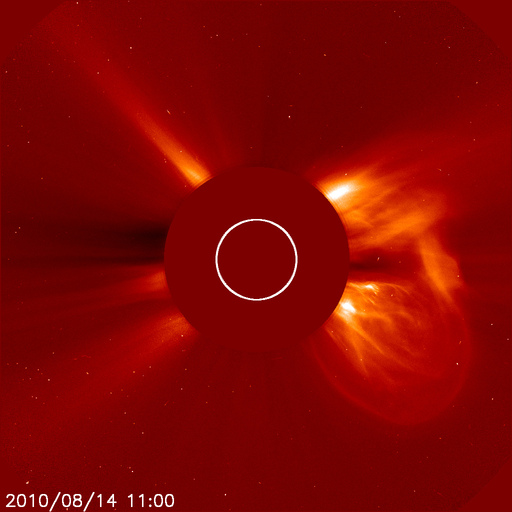}
\hspace*{-0.015\textwidth}
\includegraphics[width=0.33\textwidth,clip=]{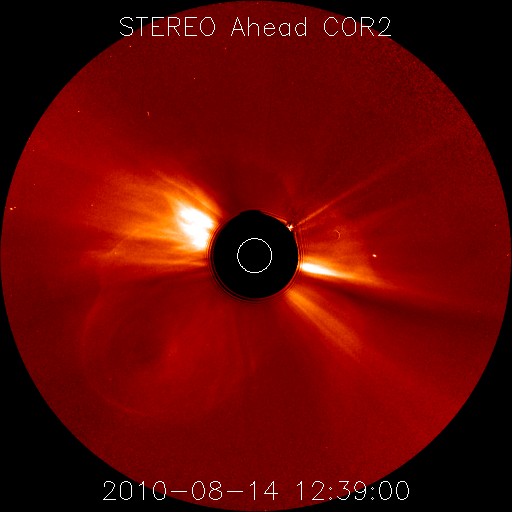}      

\includegraphics[width=0.33\textwidth,clip=]{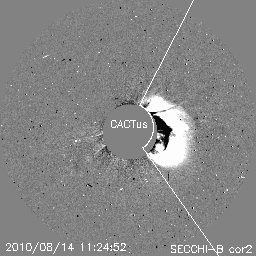}  
\hspace*{-0.015\textwidth}
\includegraphics[width=0.33\textwidth,clip=]{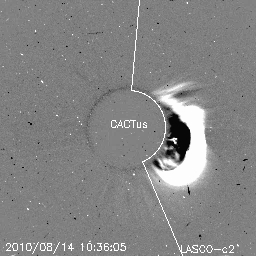}
\hspace*{-0.015\textwidth}
\includegraphics[width=0.33\textwidth,clip=]{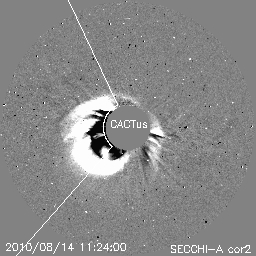}      

\caption{\textbf{Top panels:}  Observations of the CME associated with the filament eruption on 2010 August 14. These observations were made by STEREO-B/COR2, SOHO/LASCO, and STEREO-A/COR2 respectively (from left to right). \textbf{Bottom panels:} CACTus detections of the coronal mass ejections based on observations from the same instruments. The white lines indicate the estimated angular width of the CME. Images taken from \url{sidc.oma.be/cactus/}.}
\label{detectiesencactus} 
\end{figure}

\begin{table}[H]
\centering
\begin{tabular}{llccc} 
\hline 
\hline
Coronagraph & Earliest detection & Angular width & Median velocity & Principal Angle \Tstrut\\
&time (UT)& (degree)& ($\textrm{km s}^{-1}$) & (degree) \Bstrut\\
\hline
STEREO-B/COR2 & 2010/08/14 10:24 & $118~(II)$ & $641~\pm~221$ & $274$ \Tstrut\Bstrut \\
\hline
SOHO/LASCO\ & 2010/08/14 10:12 & $152~(II)$ & $657~\pm~223$ & $279$ \Tstrut\Bstrut\\
\hline
STEREO-A/COR2 & 2010/08/14 09:54 & $114~(II)$ & $568~\pm~184$ & $82$\Tstrut\Bstrut\\
\end{tabular}
\caption{Parameters for the CACTus detections of the CME in coronagraph data from the SOHO and STEREO spacecraft. Data reproduced from \url{sidc.oma.be/cactus/}. For all three detections, the angular width was larger than $90\degree$, and thus the CME was classified as a Type II halo CME.  }
\label{cactusdetections}
\end{table}

STEREO-A/COR2 observations show the first signs of this halo CME in the white light coronagraph images around 10:24~UT in the south-east. Its counterpart instrument on STEREO-B made the first observation of the CME at roughly the same time (10:39~UT). From this viewpoint, the CME was directed towards the west. LASCO, the coronagraph on-board SOHO \citep{Brueckner1995}, detected this event even earlier, at 10:12~UT, and from this point-of-view it was also directed towards the west. Some observations by these coronagraphs, made when the CME had fully come into view, are shown in the top row of Figure~\ref{detectiesencactus}. The bottom row images depict the detections of this coronal mass ejection that were made by CACTus. CACTus also determines a number of parameter values for each detection, these are listed in Table~\ref{cactusdetections} for each of the coronagraphs that observed the CME. The parameters based on observations by the different coronagraphs agree rather well. Due to the data processing that the CACTus module performs before detecting the CMEs, the CME start times may differ somewhat from those we derived from the plain coronagraph images. Note that, as the bottom right panel in Figure~\ref{detectiesencactus} clearly shows, the angular width for this CME was underestimated (by approx. $43\degree$) in the STEREO-A data because only part of the full halo CME was detected automatically by CACTus. Additionally, the CME velocity estimates in the CACTus catalog are plane-of-the-sky measurements and thus underestimate the true velocity of the CME. Using three-dimensional information, we obtain a better estimate of the CME velocity in Section~\ref{recon_section}.

\begin{figure}  
\centering
\includegraphics[width=0.32\textheight]{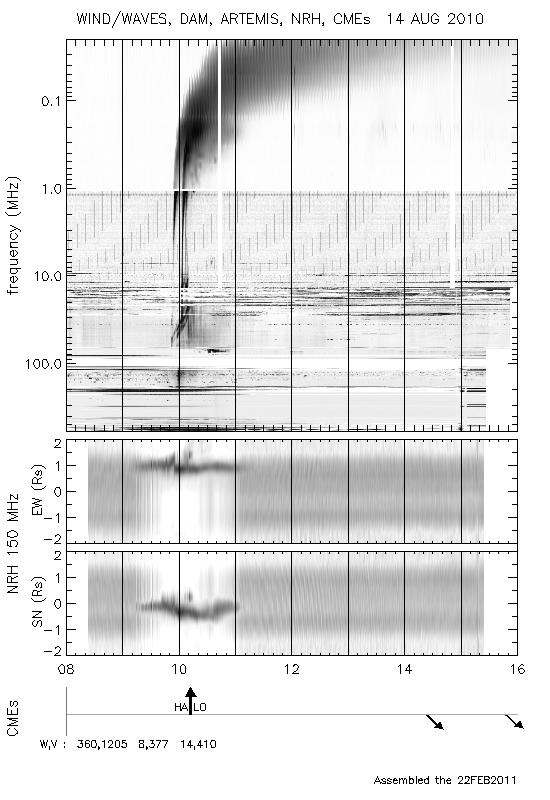}
\hspace*{0.015\textwidth}
\includegraphics[width=0.32\textheight]{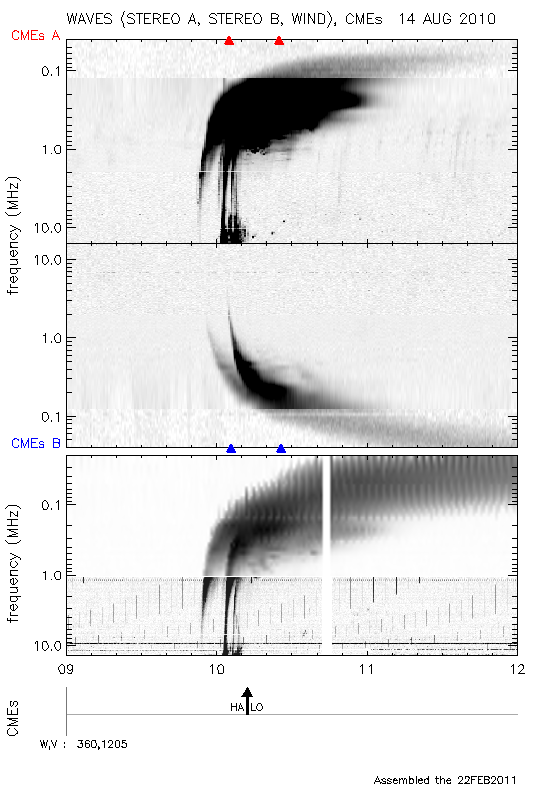}
\caption{\textbf{Left Panel:} Composite radio image combining data from the WIND spacecraft and various spectrographs, showing type II and type III radio bursts. Below are the locations of the radio sources and the reported times of associated CMEs. \textbf{Right Panel:} Radio signatures measured by the STEREO spacecraft. STEREO-A measured the strongest signatures, due to the fact that the shock propagated in the direction of that spacecraft. Images taken from \url{http://radio-monitoring.obspm.fr/}.}
\label{radio} 
\end{figure}

\subsubsection{Radio Signatures}

The Observatoire de Paris maintains a radio survey in collaboration with the Artemis team, the Universities of Athens and Ioanina and the Naval Research Laboratory. These institutes provide daily surveys which combine radio spectra covering a large frequency range from various observing sites in France, China, Australia and Greece (\url{http://radio-monitoring.obspm.fr/}). These spectra are further complemented by spectrograph information from the WIND \citep{Bougeret1995} and STEREO missions, Nan\c{c}ay Radioheliograph data and information on CME occurrence from SOHO and STEREO. 

Their data demonstrate clear \textit{radio signatures} for the event on 2010 August 14, as is shown in Figure~\ref{radio}. The left panel of this figure is a composite of radio observations from ground-based telescopes and WIND/WAVES, the radio and plasma wave investigation on-board the WIND spacecraft. This image thus shows the solar radio bursts observed from the Earth's perspective. There are clear signs of a type II burst and multiple type III bursts in this graph. The fast-drifting type III bursts occur when electrons are accelerated during the onset and impulsive phase of a flare. Type II bursts originate from the propagating shock of the CME and these burst generally drift more slowly and smoothly from high to low frequencies (compared to type III events). The type II burst occurs around 09:50~UT for this event, which corresponds to the time of the eruption of the filament as observed in EUV images. The type III bursts match the impulsive and peak phases of the flare measured by GOES (09:50~UT to 10:10~UT).

The right panel of Figure~\ref{radio} shows the radio signatures measured by the SWAVES (STEREO/WAVES) instrument on-board the STEREO spacecraft (top panels), in addition to the measurements from the WIND spacecraft (bottom panel). Again type III radio bursts are clearly observed. They are stronger for STEREO-A and WIND than for STEREO-B, simply because the shock is propagating between the location of the Earth and the STEREO-A spacecraft (as we will explain in Section~\ref{recon_section}). 

\subsubsection{Interplanetary Effects}

\begin{figure}  
\centering  
\includegraphics[height=0.23\textheight]{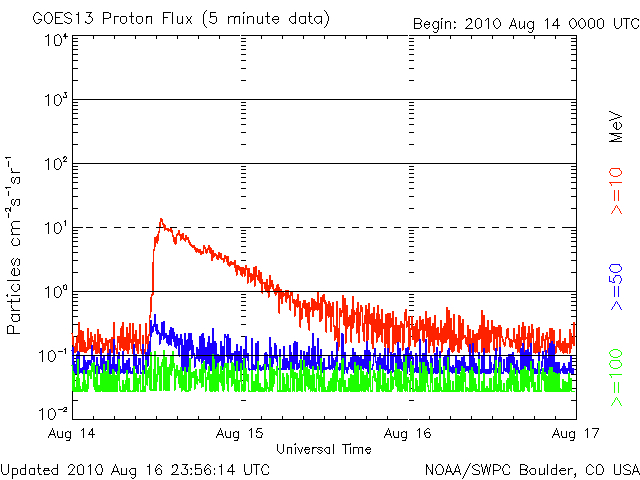}
\hspace*{0.015\textwidth}
\includegraphics[height=0.23\textheight]{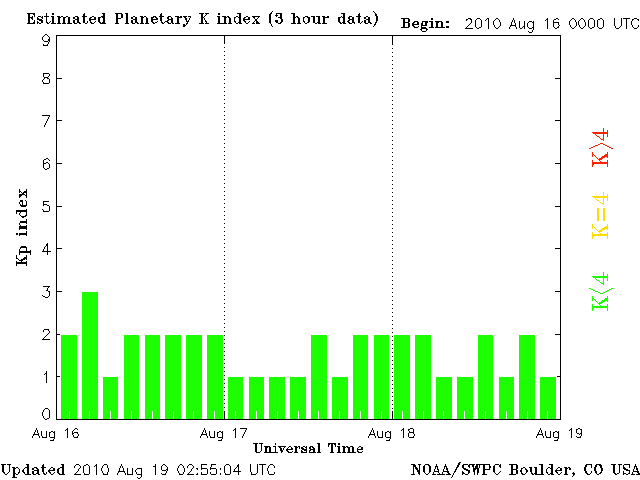}
\caption{\textbf{Left Panel:} Proton fluxes measured by the GOES spacecraft, illustrating a clear jump in the measurements of the $>10$~MeV protons, associated with the 2010 August 14 eruption. \textbf{Right Panel:} Planetary K index measured for the days after the eruption. No strong disturbance of the magnetic field is measured, resulting in quiet levels for the $K_p$ index ($K_p < 4$). Images provided by NOAA at \url{ftp://ftp.swpc.noaa.gov/pub/warehouse/}.}
\label{protons_kp} 
\end{figure}

Figure~\ref{protons_kp} (left panel) clearly shows a strong increase in the proton flux measured by the GOES spacecraft starting at 12:30 UT and reaching a maximum at 12:45 UT. The strongest peak was measured for protons with energies $>10$~MeV. The curve barely crossed the threshold of 10 proton flux units, which corresponds to a minor solar radiation storm of type S1. By the end of the next day, August 15, the proton levels had decreased back to normal levels. An S1 radiation storm is not expected to cause any problems with satellite operations, but minor impact on HF radio communication in the polar regions is possible. Nevertheless, some white streaks were visible in the SOHO/LASCO C3 images as a result of energetic particles hitting the coronagraph detector. 

\begin{figure}
\centering
\includegraphics[height=0.3\textheight]{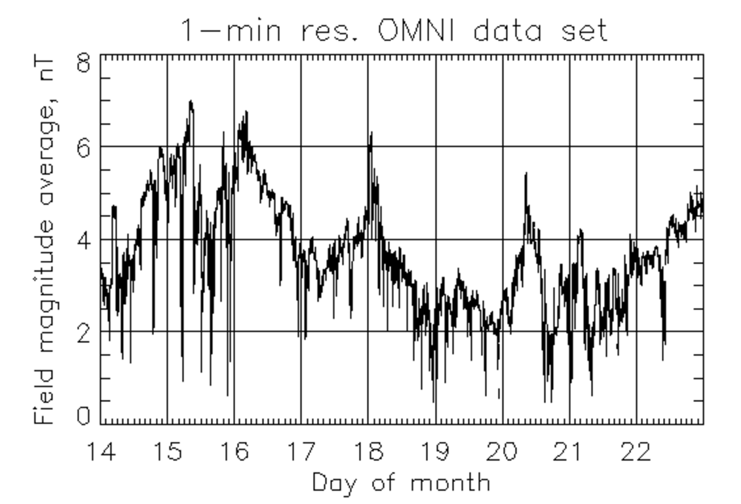} 

\includegraphics[height=0.3\textheight]{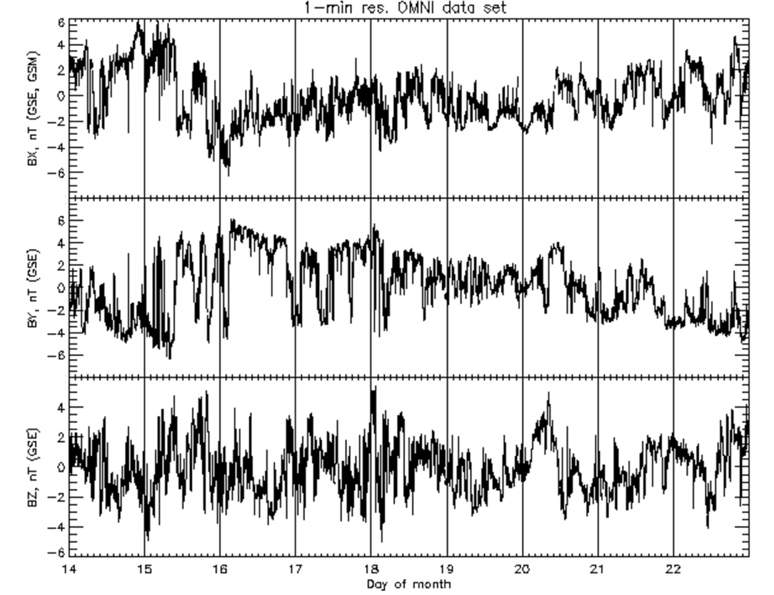} 

\includegraphics[height=0.3\textheight]{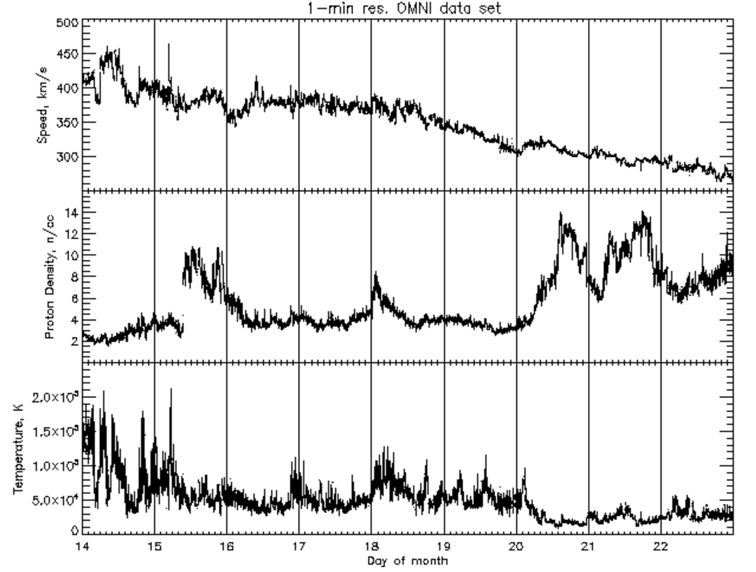} 
\caption{In-situ data from the ACE satellite measuring the conditions of the solar wind close to Earth. Plotted are (from top to bottom) the total strength of the magnetic field, the $B_x$, $B_y$ and $B_z$ component of this magnetic field, the solar wind speed, the proton density and the temperature. No clear signs of the arrival of an ICME are visible in the days after 2010 August 14. Images produced with \url{http://omniweb.gsfc.nasa.gov/form/omni_min.html} .}
\label{ace}
\end{figure}

\begin{figure}       
\centering
\includegraphics[width=0.87\textwidth]{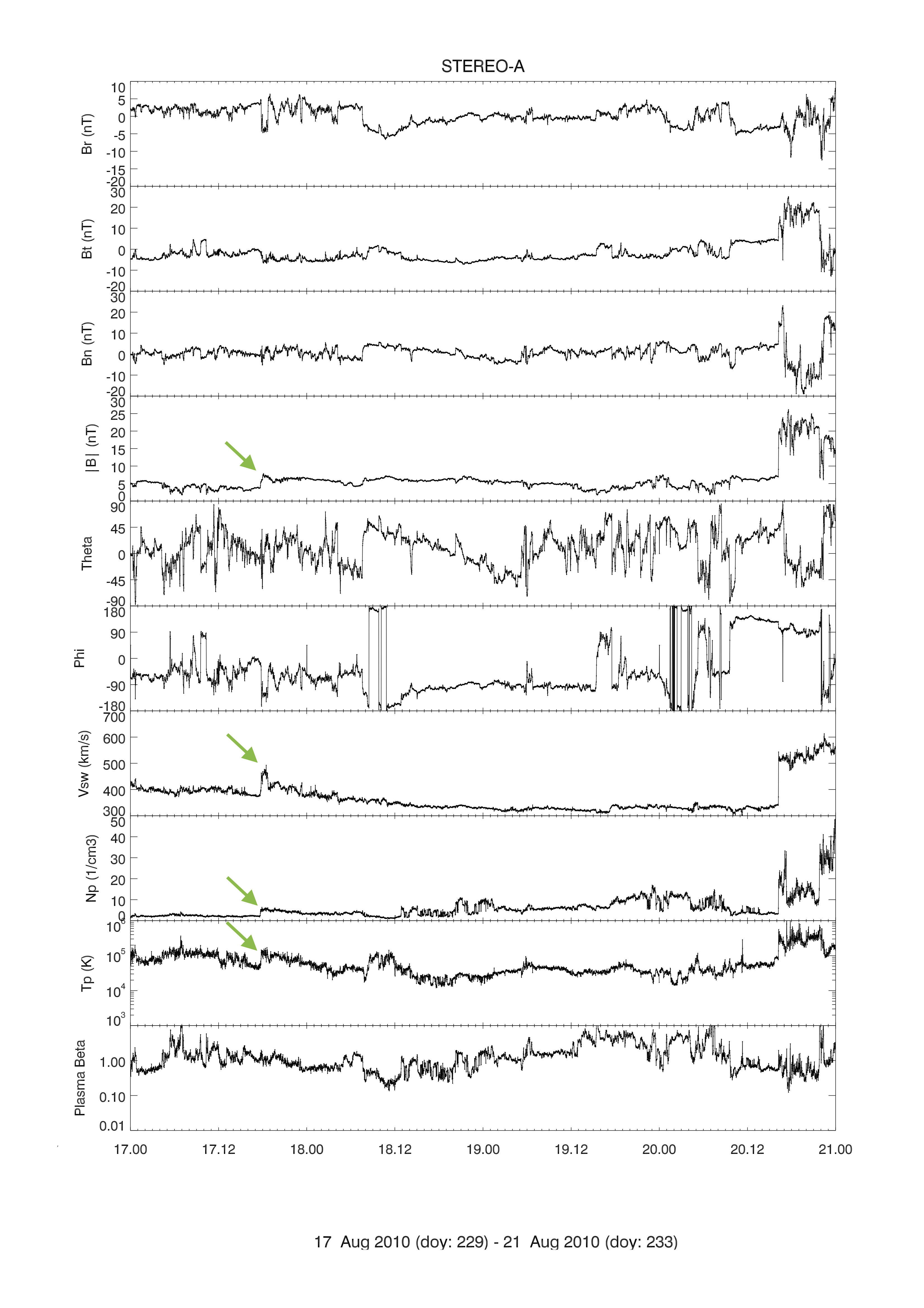}
\caption{Measurements of the solar wind obtained by the in-situ instruments IMPACT and PLASTIC on-board STEREO-A. In the afternoon of 2010 August 17, a weak shock is visible in these observations of the solar wind speed ($v_{SW}$), the proton temperature ($T_p$), the total magnetic field strength ($|B|$) and the proton density ($N_p$), indicated by the green arrows.}
\label{STA}
\end{figure}

The in-situ data recorded in the days after the eruption do not show the arrival of the \textit{interplanetary CME} (ICME) at Earth (Figure~\ref{ace}). Consequently, also the planetary K-index, a measure indicating how strongly the Earth's magnetic field is disturbed by space weather events, was at most $K_p=3$ (Figure~\ref{protons_kp}, right panel) in the days after the eruption, corresponding to quiet conditions and confirming the absence of a geomagnetical storm. 

Yet, the ICME did arrive at the STEREO-A spacecraft and can be observed in the in-situ measurements shown in Figure~\ref{STA}. This plot shows a moderate shock in the late afternoon of 2010 August 17 in the measurements of the total magnetic field strength, the solar wind velocity, the proton density and the proton temperature (indicated by the green arrows). Together with the decreasing velocity profile and the relatively low temperature, these are clear signs of the passing of an ICME. This shock seems rather small compared to the high velocity we derived for the CME (see Section~\ref{recon_section}). However, it is likely that STEREO-A only measured the flank of the CME, as it propagated between STEREO-A and Earth. The 3D reconstructions reported on below in Section~\ref{recon_section} place the source location $20\degree$ east of the central meridian as seen from the STEREO-A point-of-view. 

The STEREO-B spacecraft measurements did not show any indications of the passing of a magnetic cloud, which is to be expected as the CME was mostly directed towards STEREO-A, and thus propagating away from the STEREO-B spacecraft.

\subsection{Discussion}

A variety of solar and interplanetary signatures was associated to the filament eruption on 2010 August 14. Together, they allow to analyse this event in detail. The first clear sign of the eruption is the rise of the southern part of the filament connecting both active regions. This rising motion destroys the equilibrium of the filament and the northern part is dragged along. The filament eruption is accompanied by a mild (C4) solar flare, an EUV wave and radio signatures. As the filament detaches from the solar surface a post-eruptive arcade forms and a CME is observed in coronagraph images. Three days later, the ICME arrival is observed in STEREO-A data, but no geomagnetic impact was measured at Earth.

While the filament erupts, it seems to unwind as it is launched into space, which can be seen in SDO/AIA images between 09:15 UT and 10:15 UT. Similarly to what is described by \cite{Vourlidas2011}, this apparent rotation may be caused by the earlier disconnection of the southern end of the filament and could be studied further using 3D reconstructions. For example, also \cite{Su2013} and \cite{Bemporad2011} used three-dimensional modelling to study the rotation of an erupting filament.  

In addition to the fact that this filament unwinding in itself deserves further study, we point out that it may also significantly change the flux rope type of the CME \citep{Huttunen2005} which defines its geomagnetic impact. For example, \cite{Kay2016} and \cite{Isavnin2014} studied the impact of CME deflection and rotation on the forecast of the magnetic configuration of the CME upon impact. Additionally, \cite{Lynch2009} have described a framework to determine how the orientation of the filament or polarity inversion line (PIL) in the CME source region can be related to the orientation of the CME as observed in coronagraph images, and to the ICME orientation obtained from in-situ measurements. They also point out that the tendency of a flux rope to rotate clockwise or counterclockwise depends on its magnetic configuration. These studies illustrate that in addition to the magnetic configuration of a CME flux rope being difficult to estimate at the time of eruption, CME rotation may also change this magnetic orientation while the CME travels through interplanetary space which makes the forecast of its geomagnetic impact even more challenging.

The properties of the associated EUV wave were studied by \cite{Long2011}. Using STEREO-A/EUVI 195~\AA~images, as well as various SDO/AIA passbands, the authors report initial velocities for the EUV wave ranging from 343 to 460~$\textrm{km s}^{-1}$, indicating that these measurements are highly dependent on the passband that is used. The decelerations based on the different SDO/AIA wavelengths range form $128$ to $431~\textrm{m s}^{-2}$. Using these results, the authors applied coronal seismology to estimate the quiet-corona magnetic field strength and found that it lies within the range of 1 to 2~G. 

This eruption was also accompanied by diverse radio signatures. As mentioned above, these include Type II and III bursts. Note that \cite{Tun2013}  also observed a moving type IV radio burst in Nan\c{c}ay Radioheliograph data. The cause of this type IV burst is believed to be the gyrosynchotron emission from the core of the CME, which corresponds to the erupting filament itself. Based on this observation, \cite{Tun2013} estimated the loop-top magnetic field strength along the line of sight and at $1~R_{\odot}$ above the solar surface to lie between $5~G$ and $15~G$. Also assuming gyrosynchotron emission, \cite{Bain2014} find lower magnetic field values of only several Gauss. This disagreement may be explained by the different choice of high energy cutoff that was made in these studies \citep{Bain2014}.

The short duration of an impulsive SEP event associated to a solar eruption is sometimes explained by assuming that these particles were accelerated at the flare site rather than at the CME shock front. However, based on their relative abundances and particle profiles, \cite{Cane2010} showed that SEP events form a continuum with no specific solar parameters ---such as flare or CME association--- distinguishing them from each other. The authors did find an interesting relation between the timing of the associated Type III radio bursts and the SEP profiles. Eruptions where the associated Type III bursts occurred after the impulsive phase of the flare tend to be associated with larger SEP events. When the Type III burst was observed during the impulsive phase, this indicates a rapid acceleration and escape of particles and the proton event typically shows a lower intensity and a short duration, as was the case for the 2010 August 14 event, (S1 storm, 1 day duration). 

The effects of solar energetic particles at Earth turned out to be minimal for this event, despite the strong shock that accompanied this fast CME. Indeed, the source region for this event was located close to the solar limb when the CME erupted and the CME itself was directed mostly westward as seen from the Earth (thus directed more towards the STEREO-A spacecraft, see Section~\ref{recon_section}). It is likely that most of the accelerated particles travelled on magnetic field lines that were not connected to Earth and thus did not arrive at our magnetosphere. Despite its moderate strength, this proton event received a lot of attention as it was the first proton event affecting Earth that was recorded since December 2006 and the only one observed during the year 2010, according to NOAA\footnote{\url{http://umbra.nascom.nasa.gov/SEP/}}. \cite{Bain2016} performed a detailed study on the SEP events that were observed in August 2010 and combined observations with various modelling techniques to understand the shock connectivity for and interaction between these events.

When analyzing this CME in the daily bulletin from the Belgian RWC, the forecaster on duty noted that the CME was mostly southward, but partially Earth-directed. Due to the limited amount of data available at the time of the forecast, a preliminary estimate was made: the CME upper region was expected to skim the Earth on August 17. However, as shown in Figure~\ref{ace}, the ICME never arrived at Earth. Indeed, this preliminary assessment was based on the coronagraph data available at the time of writing (that is at 12:30 UT, just a few hours after the start of the eruption). Due to projection effects it may be difficult to estimate the true trajectory of the ICME, despite the availability of images from different viewpoints. To determine the true propagation direction of a CME, three-dimensional reconstruction (as described in Section~\ref{recon_section}) and CME modelling are needed. These techniques are available to space weather forecasters, but require some work and time and are therefore not always part of their initial assessment. The detailed analysis in Section~\ref{recon_section} explains that the CME was directed more towards the STEREO-A spacecraft and therefore did not interact with the magnetic field of the Earth.

\section{Three-dimensional Reconstructions}\label{recon_section}
Three-dimensional reconstruction methods are often used to study the trajectory, kinematics and morphology of CMEs starting from their site of initiation to their propagation into interplanetary space. By exploiting the different viewpoints of various spacecraft, this type of modeling allows to mitigate projection effects and estimate the CME's evolution in interplanetary space. This information is crucial to understand how the ejected CME interacts with the surrounding solar wind and to predict whether and how the CME will affect the Earth's magnetosphere. 

We explored the trajectory of the CME associated with the eruption on 2010 August 14 through the \textsf{scc\_measure solarsoft} program. This program is based on epipolar geometry \citep{Inhester2006} and allows the user to locate the same feature on two solar images taken from different vantage points. The user first selects a feature in an image taken from the first viewpoint. The program then displays the image from the second spacecraft's vantage point with a line over-plotted that indicates the line-of-sight from the first spacecraft. The user selects a point along this line that corresponds to the same feature that was tracked in the first image. The program then computes the heliographic coordinates of this feature in three-dimensional space. 

In the case of the 2010 August 14 event, the eruptive filament was clearly visible in EUV wavelengths. Various EUV instruments (PROBA2/SWAP, STEREO-A/EUVI and SDO/AIA) observed this prominence from different angles and saw the filament rise, destabilize and twist as it erupted into space. The further propagation of this filament into interplanetary space was imaged by the coronagraphs on-board SOHO and STEREO. 

Ideally, a three-dimensional reconstruction would allow us to track the rise and acceleration profile of the erupting flux rope without projection effects, which, in turn could yield an accurate height-time diagram.  \cite{Schrijver2008} argued that by fitting such a height-time diagram with different functions it should be possible to determine which of several eruption mechanisms was likely responsible for the onset of the eruption by comparing these fits to predictions from simulations. For example, a height-time profile with a parabolic shape matches the numerical results for the breakout model \citep{Lynch2004}. The CME rising phase in case of the catastrophe model follows a power law with exponent 2.5 \citep{Priest2002}. Finally, MHD instabilities are compatible with an exponential rising phase \citep{Torok2004, Torok2005, Kliem2006}.

\begin{figure}       
\hspace*{-0.1\textwidth}
\includegraphics[height=0.8\textheight]{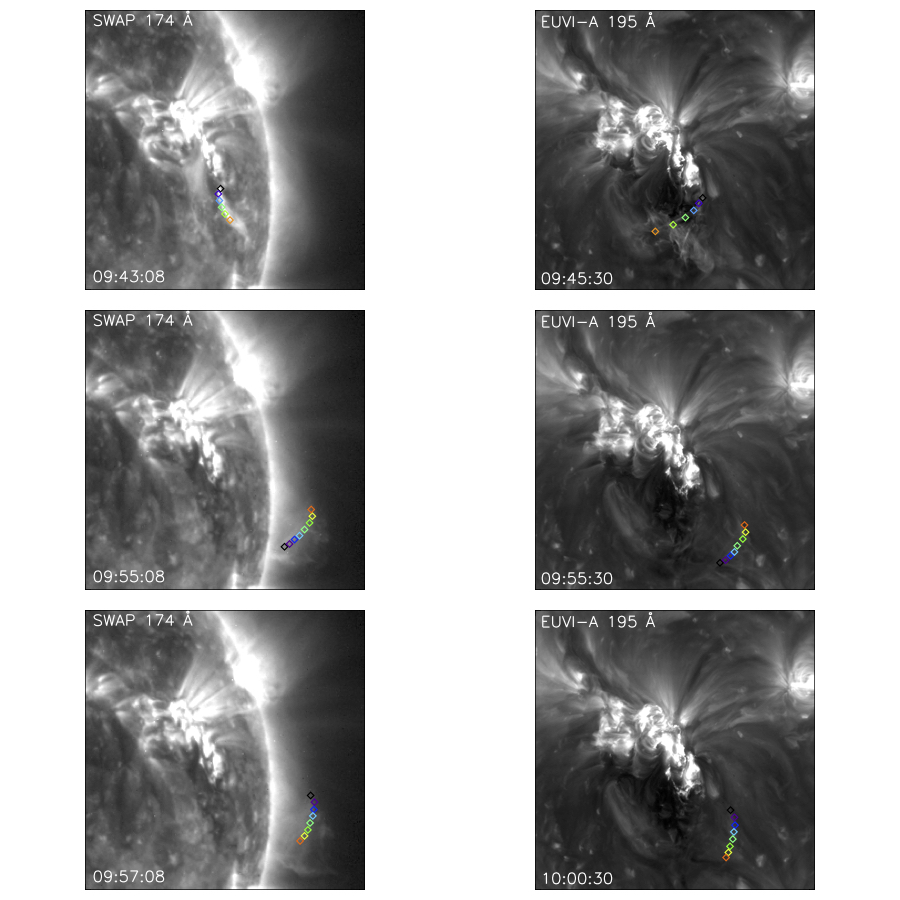}
\caption{Three-dimensional reconstruction using epipolar geometry of the erupting flux rope observed during the 2010 August 14 event. For this reconstruction, we used PROBA2/SWAP 171~\AA\ and STEREO-A/EUVI~193~\AA\ images, matching in time as closely as possible.}
\label{blob}
\end{figure}

We first reconstructed the entire erupting filament using PROBA2/SWAP and STEREO-A/EUVI images. The results are shown in Figure~\ref{blob}. This reconstruction showed that the filament erupts close to the equator at an average longitude of $65\degree$. Unfortunately, there were insufficient pairs of SWAP and EUVI-A images where the feature was clearly visible and the images well-matched in time. Thus only a few points could be measured, not enough to create a useful height-time diagram.

\begin{figure}       
\centering
\includegraphics[width=1.0\textwidth]{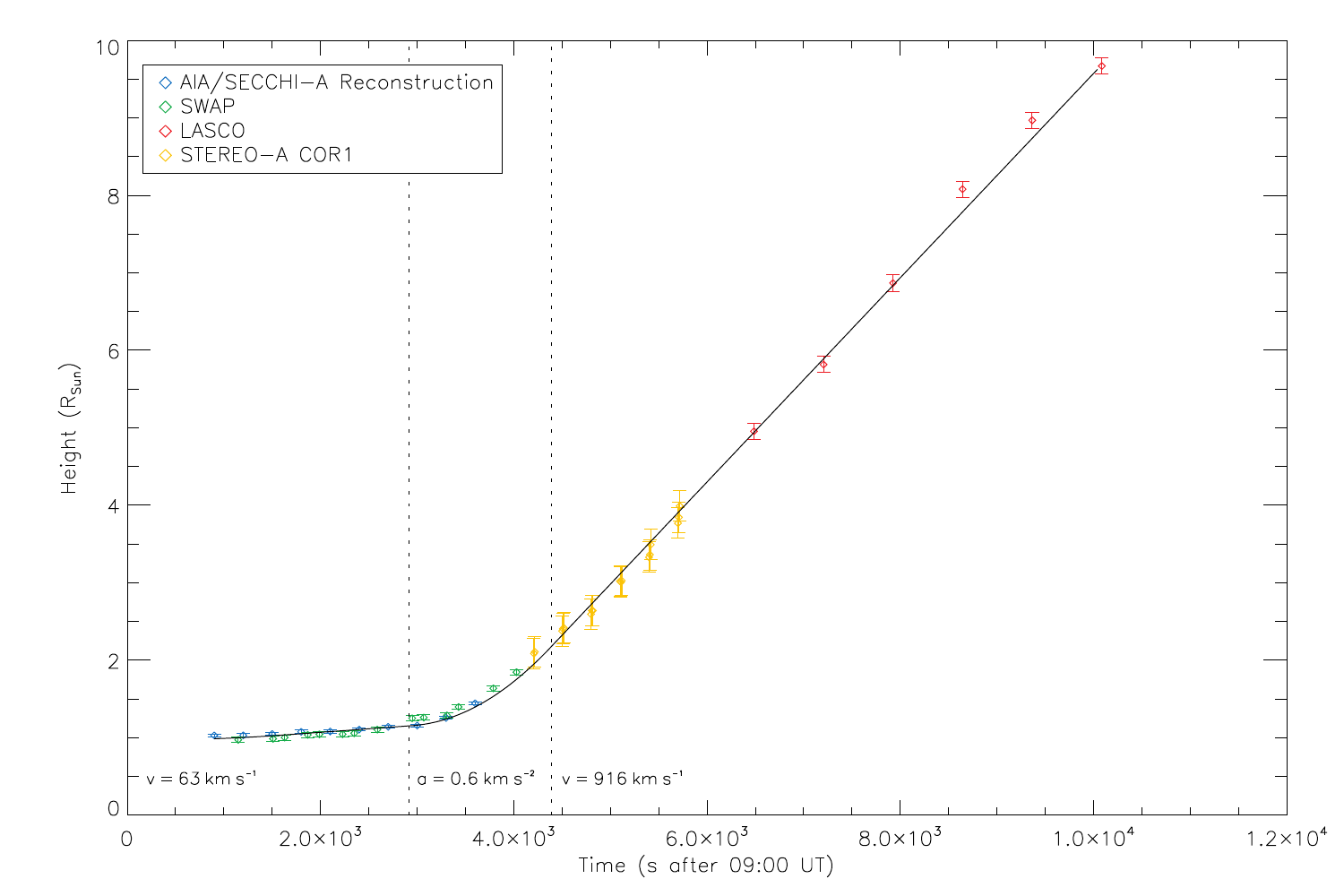}
\caption{Height-time diagram for the CME on 2010 August 14, combining measurements made using observations by different EUV imagers and coronagraphs. The measurement errors were obtained by remeasuring the position of the flux rope several times in sample images. We then applied error propagation to the standard deviation of these measurements to obtain the error bars shown here. The COR1 datapoints (in yellow) show the largest error bars because it was harder to define the exact position of the flux rope in those images.}
\label{reconstruction}
\end{figure}

We therefore expanded our dataset with the images from SDO/AIA and obtained the 3D reconstruction of the center of the bright front by combining SDO/AIA and STEREO-A/EUVI data. The result is shown in the height-time plot in Figure~\ref{reconstruction}. Because AIA has a limited field-of-view, these fitted points revealed the location of the eruption only in the very low corona. To extend the trajectory, we measured the plane-of-sky height of the feature in SWAP images, which have a large field-of-view. We then deprojected these measurements for the true propagation angle by assuming the erupting structure was traveling largely radially in the same direction as the three-dimensional reconstructions had indicated. Since the locations we obtained from our earlier reconstructions revealed points between $\mathrm{55\degree~and~65}\degree$ longitude, we assumed a propagation angle of $60\degree$ for the centre of the bright front (as seen from the Earth), which produced good agreement with the reconstructed trajectory using AIA and EUVI-A images. The resulting deprojected points are also shown in Figure~\ref{reconstruction}. 

To extend our plot to larger heights, we also analyzed the propagation of the eruption using coronagraphic instruments. First we tracked the eruption using images from SOHO/LASCO. Since SOHO views the Sun from roughly the same viewing angle as Earth, we applied the same deprojection correction to the plane-of-sky heights we measured using LASCO as we did for the measurements from SWAP. These corrected measurements appeared to align smoothly with the trajectory measured in SWAP images, but because the LASCO occulting disk blocks our view of the corona to relatively large heights above the surface, a gap remained between the SWAP-derived heights and the LASCO-derived heights.

To fill this gap, we turned to the COR~1 coronagraph on STEREO-A. COR~1 reveals the corona to much lower heights, low enough to produce data that nearly overlapped with SWAP observations. However, since the separation angle between the feature we were tracking and STEREO-A was only about $20\degree$, during the whole early part of the event the center of the erupting prominence was traveling almost directly towards the COR~1 coronagraph and was thus largely obscured behind its occulting disk until it reached greater heights. Thus, after tracking the eruption in the plane of the sky, we could no longer use the same deprojection technique that we employed on the SWAP and LASCO data. Instead, we assumed that as it reached larger heights the flux rope expanded, so its easternmost edge was traveling radially away from the Sun along with the rest of the CME, but at a much more eastward longitude, yielding a larger separation angle with COR~1.

Because we lacked another set of co-temporal coronagraphic images to use for three-dimensional reconstructions, we instead estimated the appropriate projection angle for this eastern CME edge by assuming that the deprojected COR~1 measurements should extend the trajectory we calculated for lower heights using AIA and SWAP. The plane-of-sky COR~1 measurements revealed an essentially linear trajectory in the height-time diagram, so we extrapolated backwards to the time of the largest height measurement we obtained with SWAP and determined the projection angle necessary to produce a point that matched the SWAP measurements at that time. This method yielded a longitude of roughly $40\degree$. This is roughly $20\degree$ east of the center of the eruption, not an implausible extent for a large, rapidly expanding CME. Because all of these independently deprojected points fit together to yield an essentially smooth trajectory in the plot with few outliers, we consider this validation that we have captured --- at least very roughly --- the dynamics of the eruption's onset and early propagation through the corona.

The resulting height-time diagram in Figure~\ref{reconstruction} clearly shows three different regimes: while the flux rope initially rises with a very low velocity, it gets an extra impulse and acceleration when it erupts catastrophically and propagates with a nearly constant, but very high, velocity afterwards.

We were unable to fit the points in the resulting height-time diagram with any of the functions described in \cite{Schrijver2008}, which could be because those authors confined their analysis to low heights in the corona or could be an indication that another acceleration mechanism may have come into play for this eruption. In fact, we found that the points were well-fit by a piecewise-defined function with a slow, constant-velocity rise at about $63~\mathrm{km}~\mathrm{s}^{-1}$, followed by a constant acceleration of about $0.6~\mathrm{km}~\mathrm{s}^{-2}$ for approximately 1500~s, followed by a high-speed constant-velocity propagation out of the corona at just over $900~\mathrm{km}~\mathrm{s}^{-1}$. This final velocity of the CME is compatible with what was reported in the CDAW CME catalog ($1205~\mathrm{km}~\mathrm{s}^{-1}$, see \url{http://cdaw.gsfc.nasa.gov/CME_list/}). It also fits the measurements by CACTus of the central part of the CME (see the CACTus velocity distribution on \url{http://sidc.oma.be/cactus/}), which was indeed the point that we reconstructed here. \cite{Tun2013} also reported that the CME front travels at $1204~\mathrm{km}~\mathrm{s}^{-1}$. 

In fact, the acceleration experienced by the erupting flux rope is very likely considerably stronger. The net effect of our several assumptions and the inherent error in our measurements likely acts to smooth the acceleration phase of the eruption. Inspection of the images from SDO/AIA and PROBA2/SWAP suggests that the flux rope experienced a very impulsive acceleration phase as it began to rotate and untwist after its slow rise. We discuss this impulsive acceleration in section~\ref{MHDinstability}.

\section{Initiation Mechanism}\label{initiation}

\subsection{Flux Emergence versus Flux Cancellation}

\cite{Zhang2008} studied the relationship between CME initiation and changes in the photospheric magnetic field. They found that for 60~\% of the CME source regions an increase in the large-scale magnetic flux is observed during a period of 12 hours before the eruption, while in the other 40~\% of the cases a decrease of this quantity is measured. On 2010 August 14, SDO/HMI magnetogram observations show a significant amount of flux emergence in the day and hours before the eruption, especially in the northern active region. This suggested that flux emergence played a role in triggering this eruption. 

\begin{figure}       
\centering
\includegraphics[width=0.90\textwidth]{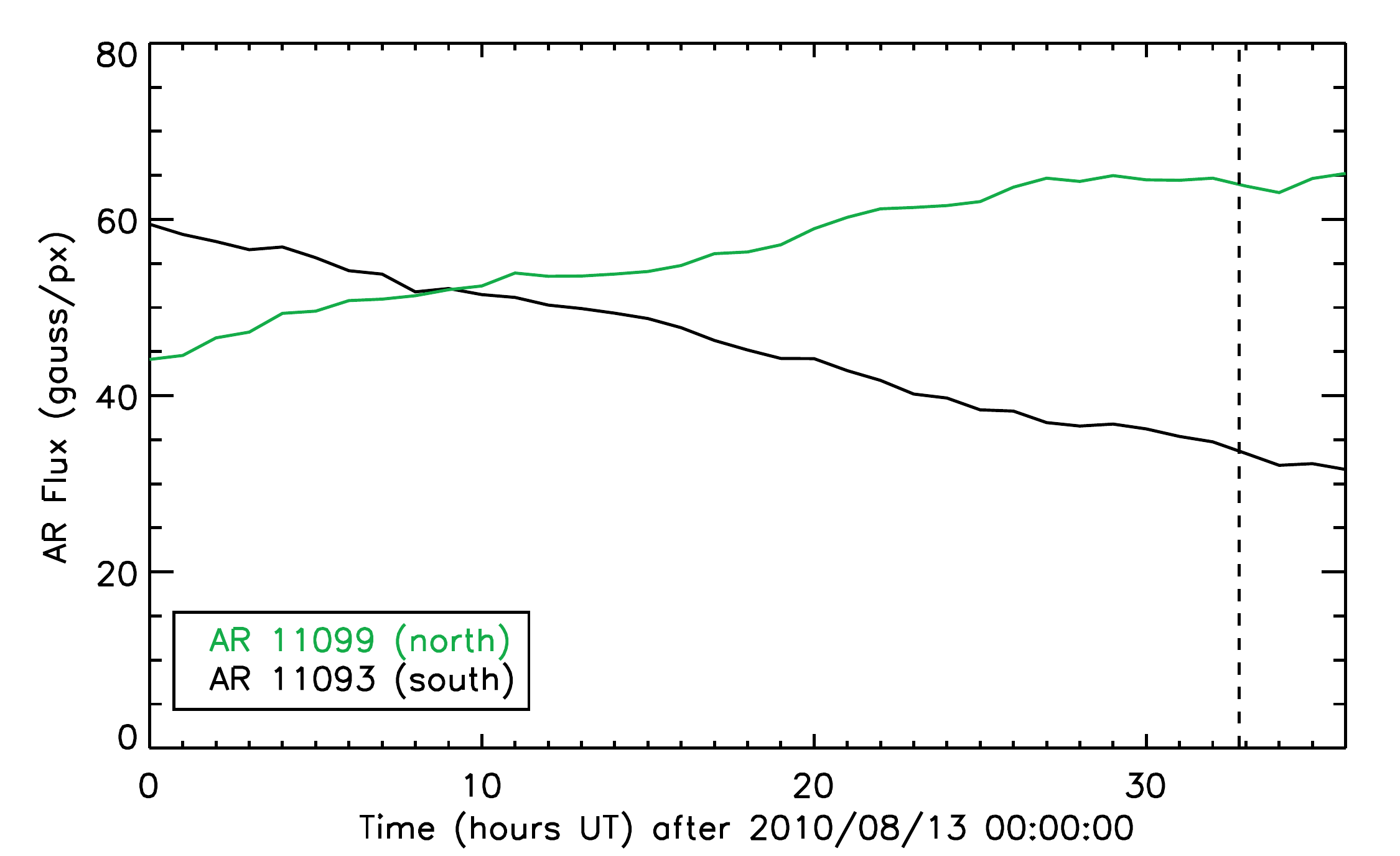}
\caption{Spatially averaged flux in the northern (green) and southern (black) active region measured by SDO/HMI over time, starting on 2010 August 13 at 00:00~UT, which is more than a day before the eruption. The dashed line indicates the approximate start time of the eruption. This figure shows a clear decrease in flux for the southern region for the entire time range. The flux in the northern region increases at first, but then levels off in the hours before the eruption. }
\label{hmi}
\end{figure}

To validate this assumption, we measured the evolution of the average flux in both active regions starting on 2010 August 13 at 00:00~UT until a few hours after the eruption on August 14. The results are shown in Figure~\ref{hmi}. These measurements revealed that there is indeed a clear increase in flux for the northern active region on the day before the event. However, the flux emergence levels off on August 14, indicating that there is no clear change in flux for this AR in the hours right before the eruption. 

On the other hand, the average flux in the southern region is steadily decreasing and continues to do so in the hours before the eruption. This suggests that the destabilization of the southern end of the filament by flux cancellation may have facilitated the eruption of the entire filament. While it is difficult to conclusively say that flux cancellation was the facilitator for this eruption, the evolution of the southern spot clearly played a role in its onset.  Additionally, this scenario closely matches the observations where the southern part of the filament is first seen to rise, dragging the entire magnetic structure with it as it erupts in an untwisting motion. 

\subsection{MHD Instability\label{MHDinstability}}

The process of initiation for this eruption may be very similar to the scenario \cite{Vemareddy2012} describe for the homologous eruption on 2010 August 7, which had the same source regions. The cancellation of magnetic flux that we observe in the southern region may be a sign of interaction of the neighboring magnetic field with the overlying magnetic field that is restraining the filament from eruption. The ubiquitous small brightenings that we observe before the eruption also fit this scenario. Although the flux cancellation in itself is insufficient to cause the eruption, tether cutting, through a series of weak reconnection events, could assist in facilitating the onset of an eruption.

Once the equilibrium in which the filament sits is sufficiently eroded, the filament is allowed to rise slowly. This early eruption phase corresponds to the left part of the height-time diagram shown in Figure~\ref{reconstruction}, with a steady but slow propagation. As the filament begins to reach larger heights, around 09:50~UT, it suddenly experiences a strong unwinding motion. This motion, in fact, shears the filamentary flux rope apart, so the southern half of the filament is accelerated extremely rapidly and the northern half becomes trapped high in the corona. These motions are very likely the result of an ideal MHD instability, probably because the inward forces on the flux rope at large heights were no longer sufficient to balance the outward forces resulting from the toroidal field of the flux rope itself. The quasi-equilibrium that facilitated its slow rise is suddenly lost and the breakup of the flux rope yields a much stronger and more rapid eruption.

As the flux rope rises further we begin to observe some signs of magnetic reconnection in its wake, including the formation of a dim post-flare arcade and the occurrence of a relatively weak solar flare. In this case, the magnetic reconnection that accompanies the event is likely a consequence of the eruption, rather than a driver of it. That the flare is relatively weak for a fast, impulsively accelerated CME suggests that the eruption's primary energy source was the ideal MHD instability that occurred when the flux rope experienced its rapid acceleration. 

Figure \ref{flare_goes} (right panel) shows an image from the Solar X-Ray Imager on the GOES-14 spacecraft, which reveals the extent of the development of the flare just before the onset of the ideal MHD instability that accelerated the CME. The image was obtained at 09:49 UT using SXI's thin polymide filter channel, which has a peak temperature response of about 7 million degrees. The bright loops in the northwest are hot, recently formed post-eruptive loops, whose formation must be linked to the onset of the eruption, since the rapid acceleration of the flux rope has not yet occurred. These loops fade gradually over time, indicating that the reconnection process that heated them has probably ceased by the time the instability sets in and accelerates the flux rope. Since there are no additional brightenings seen in SXI or in GOES X-ray irradiance measurements, we conclude that the acceleration of the flux rope was probably not associated with significant reconnection and was instead the result of an ideal MHD process.

\section{Summary and Discussion}\label{conclusion}

We studied the different aspects of an unusual solar eruption, occurring on 2010 August 14. Associated with this eruption, we observed all typical solar and space weather ingredients: a filament, a flare, an EUV wave, an EUV dimming, radio bursts, a proton event and an ICME; making it a prime example of a solar eruption. We combined data from a fleet of satellites and ground instruments to get a complete picture of this event. The most striking observational feature was not the powerful CME associated with this eruption, but its untwisting motion. Our analysis made it clear that this unwinding was most probably caused by the early destabilization of the southern end of the filament, which then only later dragged the northern part with it. 

While the initiation mechanism for this eruption was not determined with absolute certainty, we have strong indications that flux cancellation will have played an important role in the destabilisation of the southern end of the filament, which in turn led to the eruption of the entire feature. Changes in the magnetic flux were also the trigger for the eruption on August 7, originating from the same active regions. \citet{Vemareddy2012} interpreted them as signs of tether weakening. The scenario described by these authors may also be applicable to the August 14 eruption. 

We then argued that, while flux cancellation facilitated the eruption, the actual trigger was in fact an MHD instability. Only when this instability kicked in (around 09:50 UT), did the flux rope accelerate significantly. This sudden acceleration was accompanied by the formation of a shock, observed in radio observations. Afterwards, the CME propagated at a high velocity into interplanetary space.

This eruption originated from one of the first big and eruptive active regions observed during the current solar cycle. It was also associated with the first proton event recorded in nearly 4 years. Therefore, it was studied extensively by various authors, each with their own focus \citep[See, for example,][]{Long2011, Tun2013}. Understandably, the most spectacular eruptions attract the most attention and this example fits rather well in the traditional view of eruptions, where a flare and a CME occur together as a global response to a restructuring of the coronal magnetic field. 

However, in this case, the flare turned out to be much dimmer than we would expect from such a fast CME. This raises the question what is then different in eruptions where we observe a strong flare with a fast CME or, alternatively, only one of these eruptive signatures instead of both. Flares without an associated CME (confined flares) occur frequently, especially in the case of low-energy events. On the other hand, stealth CMEs, that is CMEs occurring without an observed solar flare or any other low coronal signature (such as an eruptive filament or EUV wave) are much more seldom. Based on the height-time diagram, the GOES images and our argument on the initiation mechanism in Section~\ref{initiation}, we argue that in this case the weak flare can be explained by the fact that the CME was largely accelerated by the MHD instability. The flare itself was then the result of reconnection in the wake of the CME, which was not very strong and had already stopped at the time of the flux rope acceleration.

The fact that this event was accompanied by energetic protons, a relatively fast CME and an ICME signature in STEREO-A in-situ data is interesting, considering the associated flare itself is quite weak. This emphasizes the need for careful inclusion of all possible information concerning a solar eruption in the analyses. Flare strength is not always a good indicator of the space weather risk of a CME. Space weather forecasters should therefore always consider CME properties as a whole and look at different aspects such as the presence of a flux rope, the acceleration profile, possible deflection, the presence of a shock, the source location on the Sun, etc. This is also clear from the analysis presented by \cite{Steed2012}: these authors showed that the eruption on August 7 and August 14 exhibited very similar on-disk signatures, but nevertheless had widely different space weather consequences. \cite{Webb2014}, in turn, showed that a detailed analysis of all aspects of the CME development and evolution is crucial to correctly identify the solar counterpart of an interplanetary CME. Using various models, these authors were able to confirm a faint filament eruption as the solar source of a magnetic cloud observed in-situ, despite the presence elsewhere of a bright flare, which was the first obvious suspect.

The 2010 August 14 CME originated from a compact, dense prominence that erupted rapidly and was clearly quite dynamical and structured, which argues that it was likely to be a fairly energetic event, despite the weak flare that was associated with this eruption. Due to the position of its source location with respect to ACE and STEREO-A, neither spacecraft experienced a head-on collision with the ICME. No strong geomagnetic effects were observed at Earth and STEREO-A measured only a moderate ICME. Nevertheless, had the eruption taken place just a few days before while the source regions were still facing Earth, the geomagnetic effects at Earth could have been severe (depending on the orientation of the ICMEs magnetic field). It would then also have been more difficult for space weather forecasters to analyze this eruption. Halo CMEs suffer from strong projection effects, which makes them difficult to observe head-on. Had this CME been Earth-directed and had forecasters based their assessment solely on the flare strength, this event would have been severely underestimated. It is therefore imperative to also use other diagnostics for the strength of a solar eruption. \cite{Mason2016}, for example, derived a relationship between the CME mass and velocity and the strength of the associated EUV dimming, which may in the future be used by forecasters to determine the importance of CMEs in case suitable coronagraph data is lacking. 

The difficulties with projection effects in coronagraph observations were addressed here by using the different viewpoints of the STEREO and SOHO spacecraft. Now that communication with STEREO-B is interrupted and also the STEREO-A spacecraft is ageing, our ability to perform this type of analysis is strongly hampered. Investing in a coronagraph instrument to be positioned at the L5 point would safeguard our ability to observe Earth-directed CMEs from the side and improve our forecasts of their space weather effects \citep{Webb2010}. Additionally, a second coronagraph (preferably located near the Earth) is needed to perform the multi-dimensional analyses we showed here. We currently have the LASCO coronagraph, but that instrument was launched over 20 years ago. This study of the 2010 August 14 eruption clearly illustrates the need for timely, multi-viewpoint coronagraph data to complement EUV and radio observations. Space weather forecasters need a wide range of observations of a solar eruption to complete their forecasts: one signature of eruption alone will never paint the complete picture.

\begin{acknowledgements}
The authors are grateful to Luciano Rodriguez, Marilena Mierla, Jasmina Magdalenic, Christophe Marqu\'{e} and Matthew West for valuable input. They also thank the referees for their constructive comments that allowed to improve the focus and clarity of this paper. This research was co-funded by a Supplementary Researchers Grant offered by the Belgian Federal Science Policy Office (BELSPO) in the framework of the Scientific Exploitation of PROBA2, the Inter-university Attraction Poles Programme initiated by BELSPO (IAP P7/08 CHARM), and a grant offered by the National Oceanic and Atmospheric Administration's GOES-R Visiting Scientist Program. E. D'Huys and D.B. Seaton also acknowledge support from BELSPO through the ESA-PRODEX program, grant No. 4000103240. This paper uses data from the CACTus CME and Solar Demon catalogs, generated and maintained by the SIDC at the Royal Observatory of Belgium (\url{www.sidc.be/cactus} and \url{solardemon.oma.be/}). PROBA2/SWAP is a project of the Centre Spatial de Li\`{e}ge and the Royal Observatory of Belgium funded by BELSPO. The editor thanks Nandita Srivastava and two anonymous referees for their assistance in evaluating this paper.
\end{acknowledgements}


\bibliography{aug14papers.bib}


\end{document}